\documentclass[trackchanges]{aastex701}

\usepackage{graphicx}
\usepackage{subcaption}
\usepackage{float}

\begin{document}

\title{
Revealing the high redshift host galaxy of the short GRB 061201 with JWST}

\author[0009-0005-4286-643X]{Yuhan Mao}
\email{yhmao@pmo.ac.cn}
\affiliation{Purple Mountain Observatory, Chinese Academy of Sciences, Nanjing 210023, China}
\affiliation{School of Astronomy and Space Science, University of Science and Technology of China, Hefei 230026, China}

\author[0009-0005-5074-4571]{Hanrui He}
\email{hrhe@pmo.ac.cn}
\affiliation{Purple Mountain Observatory, Chinese Academy of Sciences, Nanjing 210023, China}
\affiliation{School of Astronomy and Space Science, University of Science and Technology of China, Hefei 230026, China}

\author[0000-0002-9037-8642]{Jia Ren}
\email{renjia@pmo.ac.cn}
\affiliation{Purple Mountain Observatory, Chinese Academy of Sciences, Nanjing 210023, China}

\author[0000-0002-8385-7848]{Yun Wang}
\email{wangyun@pmo.ac.cn}
\affiliation{Purple Mountain Observatory, Chinese Academy of Sciences, Nanjing 210023, China}
\affiliation{Department of Astronomy, University of California, Berkeley, CA 94720-3411, USA}

\author[0000-0003-2915-7434]{Hao Zhou}
\email{haozhou@pmo.ac.cn}
\affiliation{Purple Mountain Observatory, Chinese Academy of Sciences, Nanjing 210023, China}
\affiliation{INAF - Osservatorio Astronomico di Brera, via E. Bianchi 46, 23807 Merate (LC), Italy}

\author[0009-0000-9352-6447]{Qiuli Wang}
\email{qlwang@pmo.ac.cn}
\affiliation{Purple Mountain Observatory, Chinese Academy of Sciences, Nanjing 210023, China}
\affiliation{School of Astronomy and Space Science, University of Science and Technology of China, Hefei 230026, China}

\author[0000-0002-9106-8718]{Yiming Zhu}
\email{ymzhu@niaot.ac.cn}
\affiliation{Nanjing Institute of Astronomical Optics $\&$ Technology, Chinese Academy of Sciences, \\Nanjing 210042, China}
\affiliation{CAS Key Laboratory of Astronomical Optics $\&$ Technology, Nanjing Institute of Astronomical Optics \& Technology, \\Nanjing 210042, China}

\author[0000-0003-4977-9724]{Zhiping Jin} 
\email{jin@pmo.ac.cn}
\affiliation{Purple Mountain Observatory, Chinese Academy of Sciences, Nanjing 210023, China}
\affiliation{School of Astronomy and Space Science, University of Science and Technology of China, Hefei 230026, China}

\author[0000-0002-9758-5476]{Daming Wei}
\email{dmwei@pmo.ac.cn}
\affiliation{Purple Mountain Observatory, Chinese Academy of Sciences, Nanjing 210023, China}
\affiliation{School of Astronomy and Space Science, University of Science and Technology of China, Hefei 230026, China}

\correspondingauthor{Zhi-Ping Jin}
\email{jin@pmo.ac.cn}

\begin{abstract}
Using deep near-infrared and optical images from JWST and HST, we identify a new host galaxy candidate for GRB\,061201. It lies at an angular separation of about $2^{\prime\prime}$ from the optical afterglow position. Photometric redshift fitting yields a best-fit redshift of $z\sim1.2$. We compare the previously proposed host at $z=0.111$ with the new candidate. The chance-coincidence probability of the new candidate is $P_{\mathrm{cc}}=0.18$, above the classical threshold of $0.1 $but consistent with a physical association given the extreme depth of JWST imaging. In contrast, evaluated with corresponding JWST observations, the previously claimed host has a lower $P_{\mathrm{cc}}=0.11$, which is driven primarily by bright-tail statistics rather than a more plausible association.
A high-redshift origin is favored by three independent lines of evidence.
First, for the $z=0.111$ scenario, the beaming-corrected energy shows that GRB\,061201 is an outlier of the Ghirlanda($E_{\rm p,i}-E_{\gamma}$) relation for short GRBs, while for the  $z=1.2$ scenario, GRB\, 061201 is well consistent with the Amati relation.  
Second, deep near-infrared observations rule out a kilonova similar to AT2017gfo at $z=0.111$.  
Third, afterglow modeling yields an AIC 
criterion of $\Delta\mathrm{AIC}=16.35$, providing strong evidence that the high-redshift scenario is favored. 
Assuming the host candidate is the actual host galaxy of GRB\,061201, the physical offset is $16.4$--$16.9$~kpc (substantially reduced from $\sim 42$~kpc) and the host stellar age is $\sim2$~Gyr, which are consistent with the host population of short GRBs. A low-redshift origin would lead to a very high binary neutron star merger rate of $\sim1400$~Gpc$^{-3}$~yr$^{-1}$, which is contradictory to the gravitational-wave constraint. We suggest that GRB\,061201 originates from a moderately high-redshift ($z\sim1.2$) host, significantly alleviating this apparent merger rate discrepancy. This case demonstrates the power of deep \textit{JWST} exposures in revealing the host galaxies of historically hostless GRBs.
\end{abstract}

\keywords{Short gamma-ray bursts (1653), Gamma-ray burst host galaxies (628), Kilonovae (2137), Photometry (1528)}

\section{Introduction}
Gamma-ray bursts (GRBs) are among the most energetic transient phenomena in the Universe \citep{1973ApJ...182L..85K, 2015PhR...561....1K}. Short-duration GRBs (sGRBs; $T_{90}<2$~s) are commonly attributed to the mergers of compact-object binaries, such as double neutron-star (NS--NS) or neutron-star--black-hole (NS--BH) systems \citep{1989Natur.340..126E, 1992ApJ...395L..83N, 2005Natur.438..994B, 2017ApJ...848L..12A}. Confirming a host galaxy is particularly important for bursts without spectroscopic redshift obtained from the afterglow: a host redshift facilitates the estimation of intrinsic energetics and provides constraints on jet opening angles, ambient density, and other physical parameters \citep{2015ApJ...815..102F,2018ApJ...857..128J,2022MNRAS.515.4890O}. 

Nevertheless, a number of sGRBs remain without an unambiguous host in deep follow-up observations and are thus labeled as ``hostless'' \citep{2010ApJ...722.1946B}. Two principal scenarios have been proposed for such bursts. First, the binary progenitor may receive a substantial natal kick at formation, causing the system to merge at a large physical offset from its birth galaxy; predicted offsets can reach tens to hundreds of kpc depending on the merger timescale and kick velocity \citep{1999MNRAS.305..763B}. Second, it is not easy to detect an intrinsically faint, high-redshift host galaxy \citep{2010ApJ...722.1946B, 2022MNRAS.515.4890O}.

Discovered by \textit{Swift}, GRB~061201 serves as an example of this long-standing ambiguity. Despite a clearly detected optical afterglow, its spectrum is faint and featureless, thus the spectroscopic redshift can not be estimated. Early deep imaging with the Very Large Telescope (VLT) failed to reveal a coincident host galaxy, leaving the burst classified as hostless for nearly two decades \citep{2022MNRAS.515.4890O}. Due to the lack of a bright, coincident underlying galaxy, GRB\,061201 has long been classified as an archetypal ``hostless'' burst. To explain its origin, two competing scenarios have been debated in the literature. In the past, the low-redshift interpretation was more widely adopted: the burst was associated with a nearby galaxy (hereafter G1) at $z=0.111$, and the large angular offset was attributed to a substantial natal kick \citep{2007A&A...474..827S, 2010ApJ...708....9F}. Alternatively, a high-redshift origin was also proposed, suggesting that the burst actually occurred within a faint, distant background galaxy directly at the afterglow position that simply evaded detection by shallower surveys.

However, this low-redshift scenario faces several physical challenges. It would require an unusually narrow jet (opening angle $\sim1^\circ$) to match the afterglow properties, leading to a beaming-corrected energy ($E_\gamma$) that falls well outside the typical Ghirlanda relation ($E_{\rm p,i}-E_\gamma$) for short GRBs. Moreover, the low-redshift origin of GRB~061201 ($z=0.111$) would lead to a high binary neutron star merger rate, which is in strong tension with gravitational-wave constraints \citep{jin2026neutron}. Consequently, some studies have has long argued for an undetected host galaxy with a higher redshift \citep{2007A&A...474..827S}.

In this paper, we present a newly detected host candidate of GRB~061201 with deep near-infrared and optical imaging from \textit{JWST} and \textit{HST}. We identify a previously unrecognized galaxy candidate (G2) located $\sim2^{\prime\prime}$ from the afterglow position. By fitting the luminosity redshift, calculating the probability of chance-coincidence, and adopting a series of independent physical diagnostic methods (including beam-corrected energy, kilonova limit, afterglow modeling, and binary merger rate constraints), we demonstrate that G2 is the most plausible host galaxy at $z\sim1.2$ and that the low-redshift hypothesis is disfavored. This offers a solution to the long-standing ambiguity of GRB~061201 and provides a self-consistent picture for its origin.

The paper is organized as follows. Section 2 summarizes the archival observations of GRB~061201 and outlines the historical ambiguity about the host galaxy. Section 3 introduces the new candidate of the host galaxy detected in \textit{JWST} and \textit{HST} images and presents our multi-band photometry. Then, we derive the photometric redshift and chance-coincidence probabilities ($P_{\rm cc}$) for both the newly discovered galaxy and the previously claimed host, and constrain the physical properties of the new candidate. In section 4, we compare the physical offset and the age of the host candidate with corresponding parameters of host galaxies of sGRBs, calculate energies of the burst with different assumed redshifts (the $E_{\rm p,i}-E_{\rm iso}$ and $E_{\rm p,i}-E_\gamma$ relations), assess kilonova limits, and illustrate the potential inconsistencies in the binary neutron star merger rate if the low-redshift origin is adopted. Section 5 presents our conclusions.

Throughout this paper, we adopt a standard flat $\Lambda$CDM cosmology with $H_0 = 69.6 \rm km \ s^{-1} \ Mpc^{-1}$, $\Omega_M = 0.286$, and $\Omega_\Lambda = 0.714$\citep{2014ApJ...794..135B}.

\section{Observations and Data Reduction}

\subsection{Swift}

Swift/BAT detected GRB~061201 on 2006 December 1 at 15:58:36 UT. Its prompt emission displays a typical short-GRB morphology, characterized by a double-peaked structure, a hard-to-soft spectral evolution, and negligible spectral lag \citep{2007A&A...474..827S,Marshall2006Report}.

Subsequent Swift/XRT observations revealed an X-ray afterglow adequately described by an absorbed power-law model. Notably, the X-ray light curve exhibits a distinct temporal break at $t_b = 2386^{+584}_{-521}$\,s, from a shallow decay ($\alpha_1 = 0.54 \pm 0.08$) to a steeper decay ($\alpha_2 = 1.90 \pm 0.15$). This break has been widely interpreted as a likely jet break \citep{2007A&A...474..827S,Marshall2006Report}.

The afterglow was detected by Swift/UVOT in the $White$, $U$, $uvw1$, $uvm2$, and $uvw2$ bands, but not detected in the $B$ and $V$ bands. 

\subsection{VLT and SOAR}

\begin{figure*}[t]
    \centering

    \subcaptionbox{R 20061202}{\includegraphics[width=120pt]{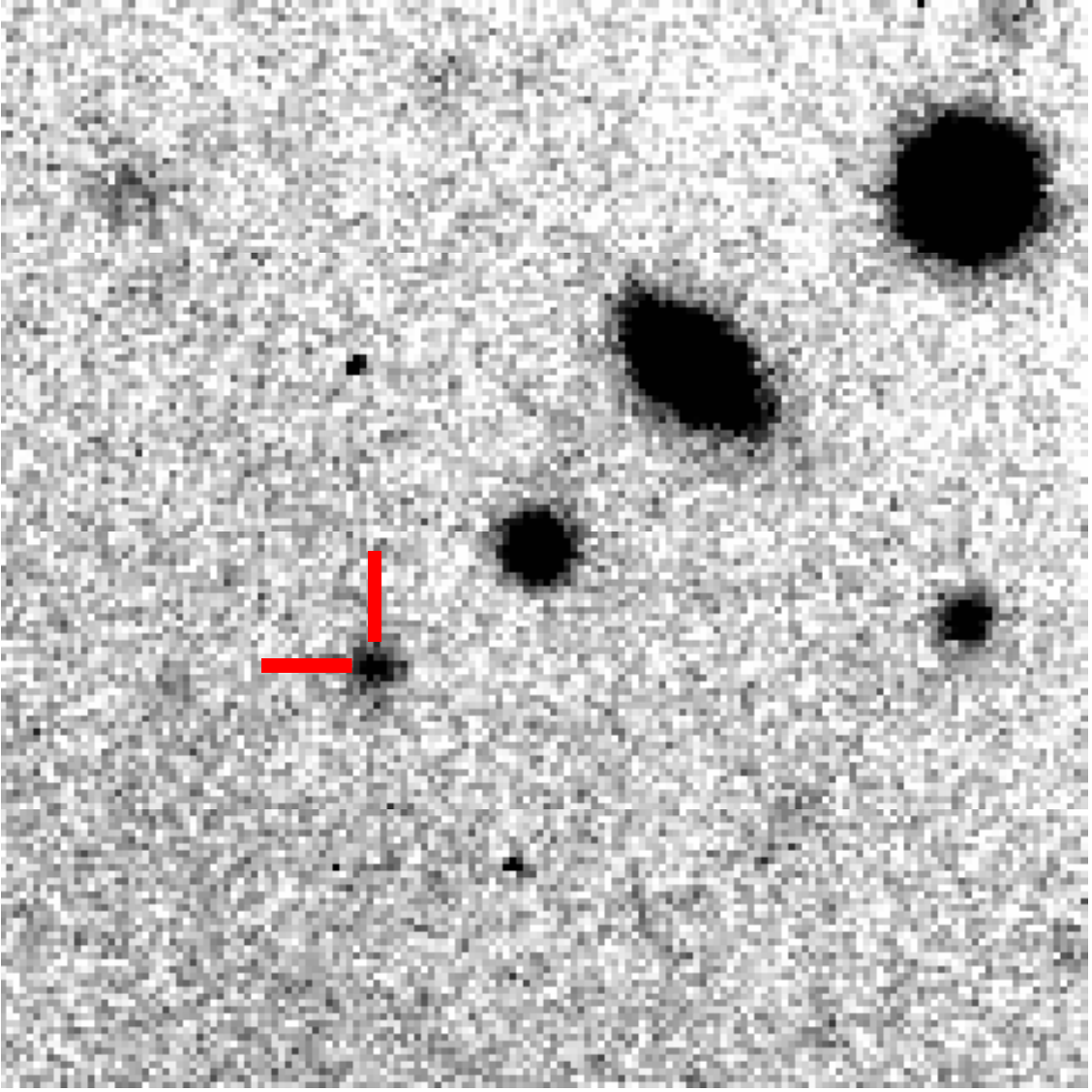}}
    \hspace{4pt}
    \subcaptionbox{I 20061202}{\includegraphics[width=120pt]{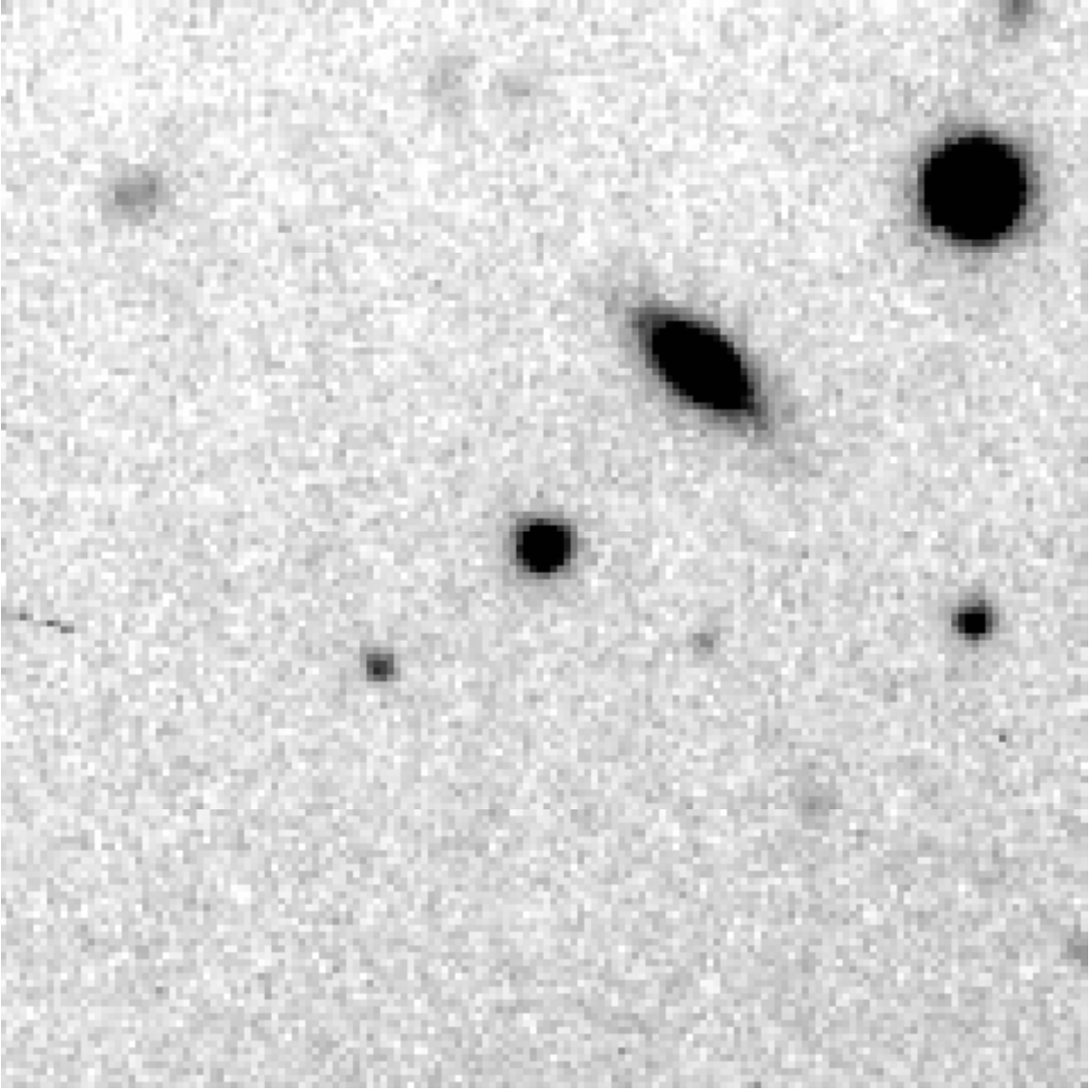}}
    \hspace{4pt}
    \subcaptionbox{J 20061202}{\includegraphics[width=120pt]{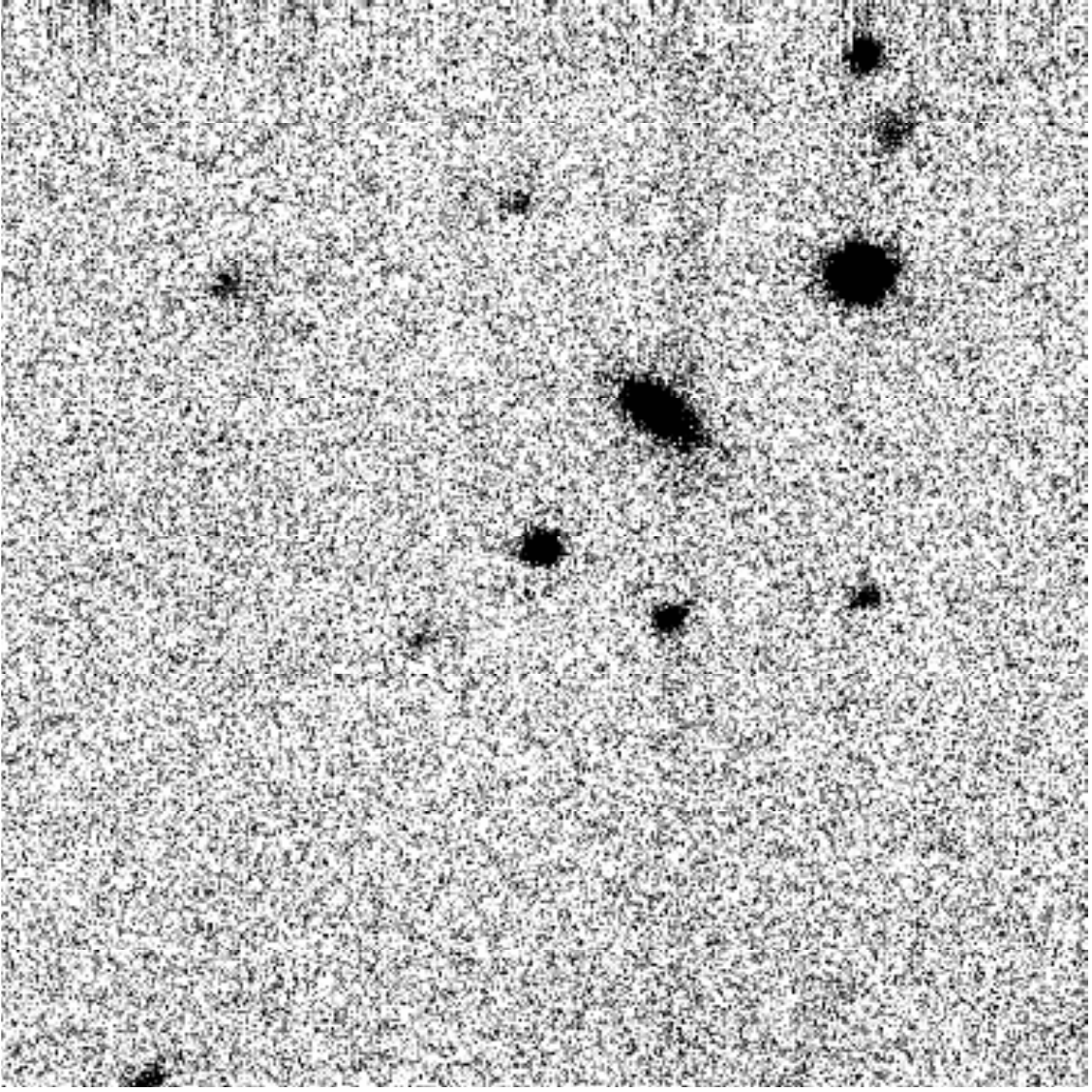}}
    \hspace{4pt}
    \subcaptionbox{K 20061202}{\includegraphics[width=120pt]{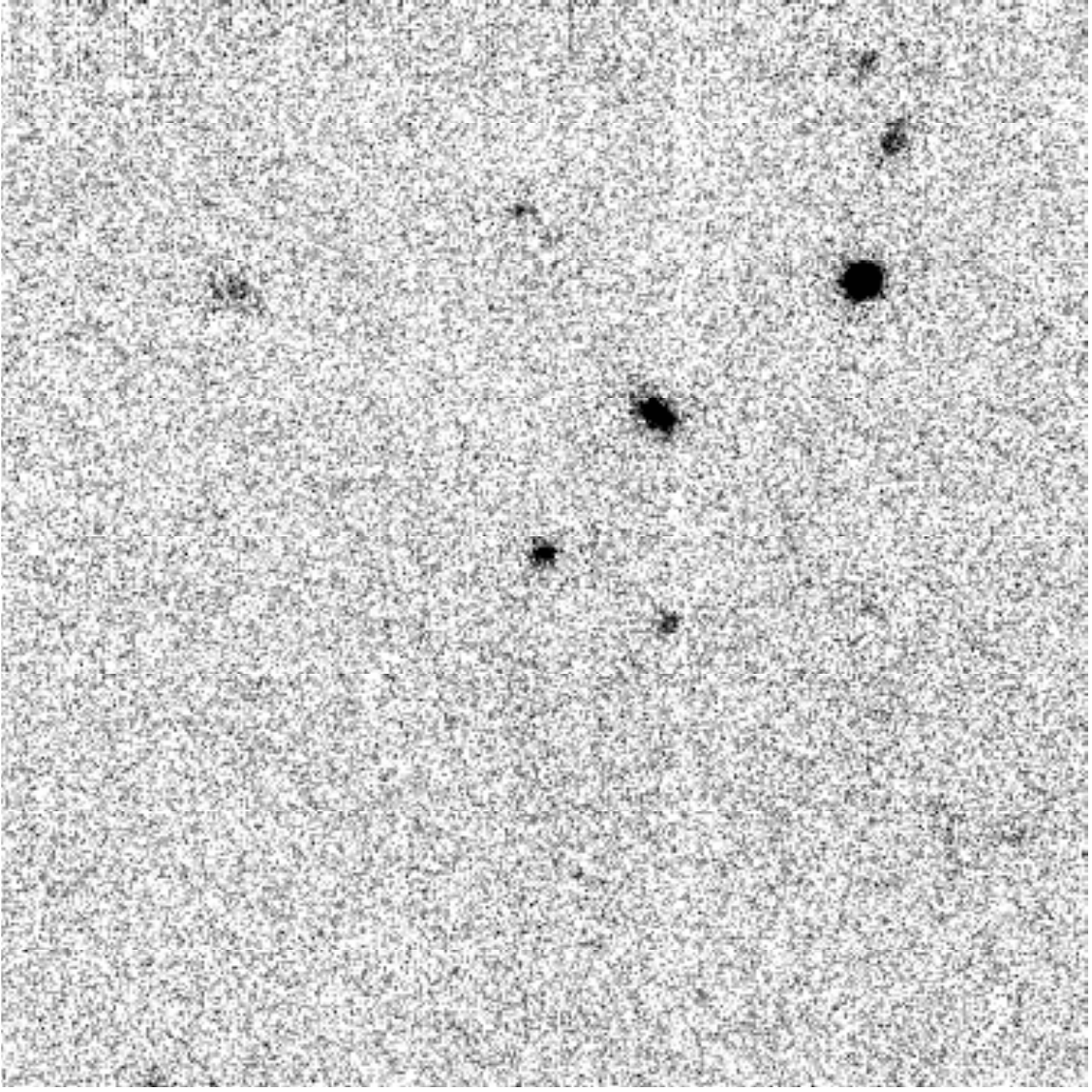}}

    \vspace{4pt}
    
    \subcaptionbox{R 20070522}{\includegraphics[width=120pt]{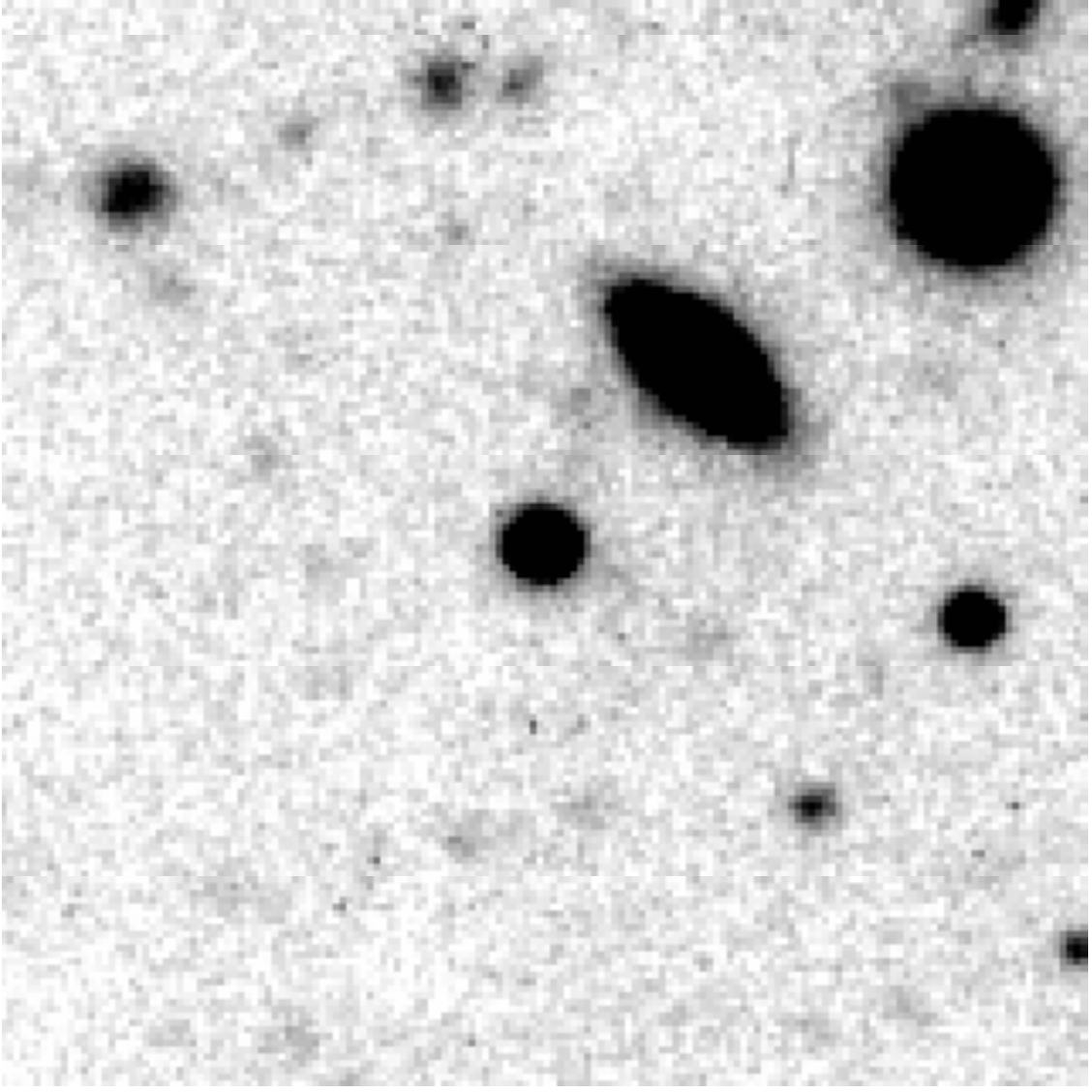}}
    \hspace{4pt}
    \subcaptionbox{I 20061218}{\includegraphics[width=120pt]{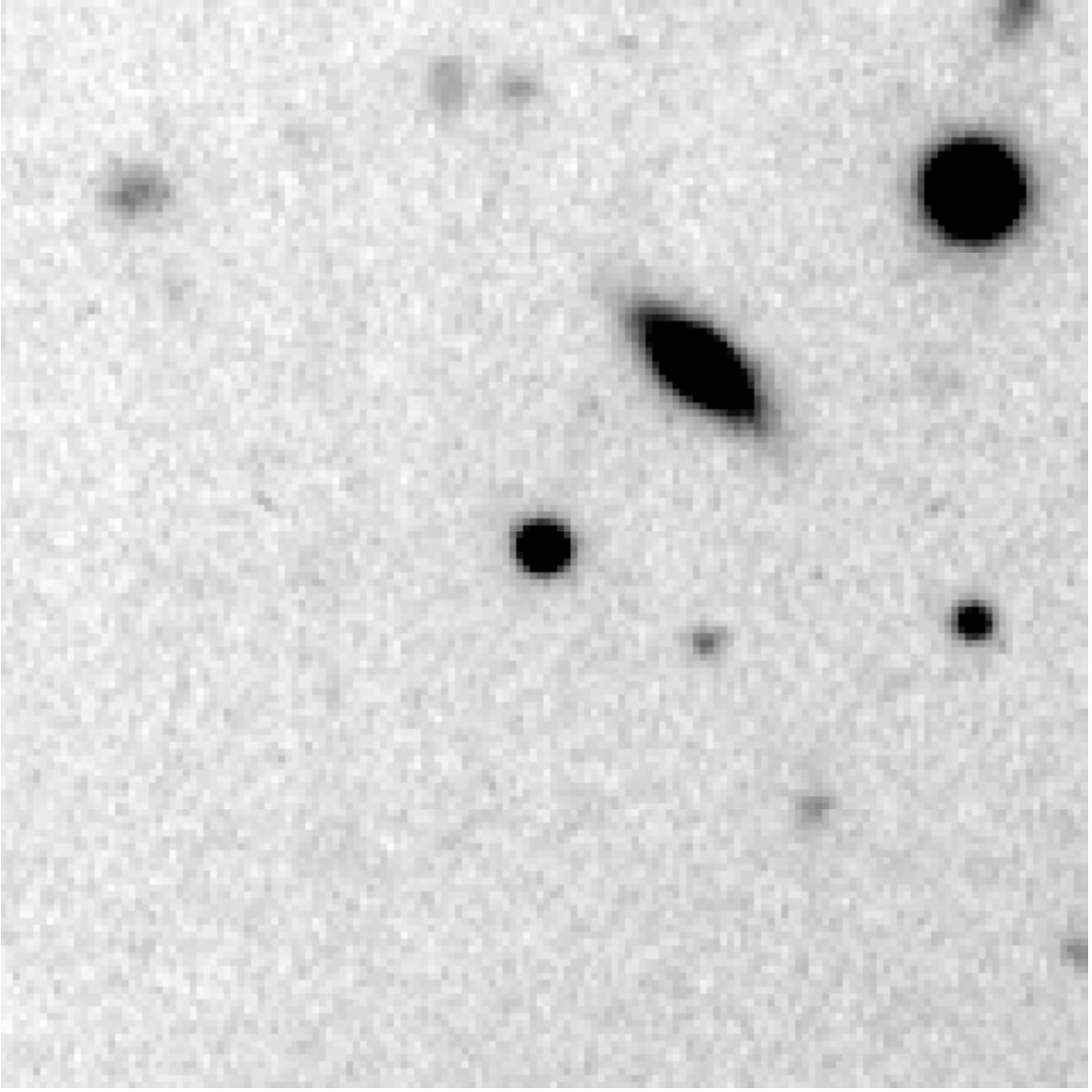}}
    \hspace{4pt}
    \subcaptionbox{J 20061203}{\includegraphics[width=120pt]{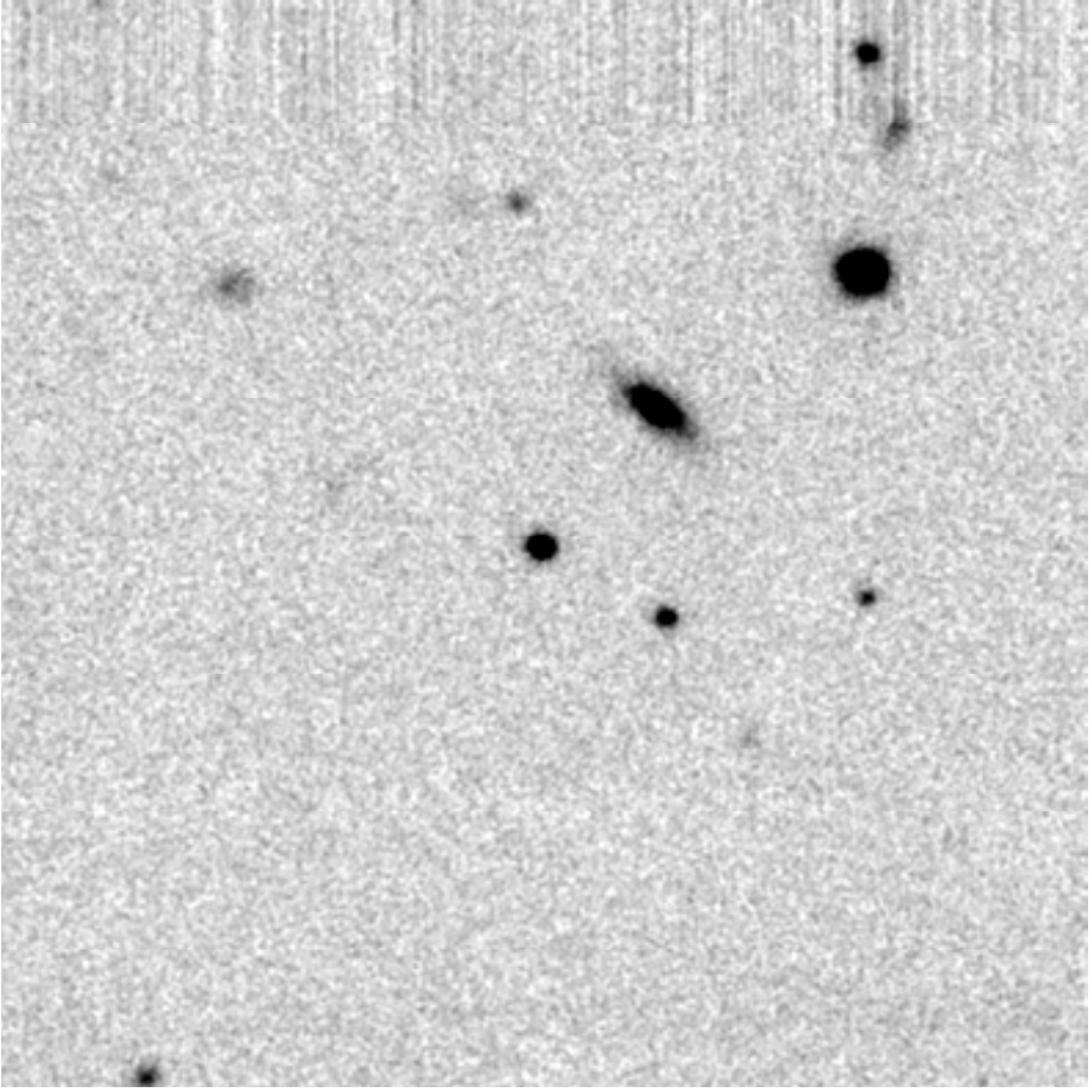}}
    \hspace{4pt}
    \makebox[120pt]{} 
    
    \caption{VLT and SOAR observations of GRB~061201. The red line segments point to the position of the afterglow. The images are oriented with North at the top and East to the left.}
    \label{fig:GRB061201VLTSOAR}
\end{figure*}

Ground follow-up observations with the VLT at 8.6\,hr post-trigger identified the fading optical afterglow with $R = 23.05 \pm 0.12$ and $I = 22.36 \pm 0.08$. 
The afterglow was also detected in the $J$ band by SOAR, yielding a magnitude of $21.51 \pm 0.13$ at $\sim 11.5$\,hr post-trigger. 
However, the $K$-band observation conducted by SOAR did not detect the afterglow, and the $3\sigma$ upper limit was $19.8$\,mag. 
Subsequent deeper $J$-band exposures with SOAR also did not detect the afterglow, and no underlying host galaxy was found at the afterglow position in any VLT or SOAR images (see Figure~\ref{fig:GRB061201VLTSOAR}).

Earlier studies\citep{2007A&A...474..827S,2010ApJ...708....9F,2010MNRAS.406.1248S} proposed a bright nearby galaxy (G1) (see Figure~\ref{VLT_JWST}(a)) as a potential host, with a measured spectroscopic redshift of $z=0.111$ based on prominent [O~II] and H$\alpha$ emission lines. The angular separation of $\sim 17^{\prime\prime}$ between the galaxy center and the afterglow position corresponds to a projected physical distance of $\sim 42$\,kpc at this redshift. Assuming a merger timescale of 0.1--3\,Gyr, this offset requires a natal kick velocity of $10$--$400 \ {\rm km \ s^{-1}}$, which is kinematically feasible. Another suggested association involves the galaxy cluster ACO S 995 at $z=0.0865$; however, the implied projected offset ($>0.9$\,Mpc) and exceptionally high required kick velocity ($>10^3 \ {\rm km \ s^{-1}}$) make this scenario highly disfavored.

Furthermore, a low-redshift assumption introduces a significant energetic tension. If GRB~061201 originated at $z = 0.111$, its isotropic-equivalent energy ($E_{\rm iso} \sim 10^{50}$\,erg) appears consistent with the isotropic-equivalent energy of the sGRB population. However, the beaming-corrected energy is $E_\gamma \sim 10^{46}$\,erg, which is significantly smaller than observed value ($E_\gamma \sim 10^{49}$--$10^{51}$\,erg) of other sGRBs. Such an extreme sub-energetic nature has not been robustly observed among securely classified sGRBs. A high-redshift origin ($z \sim 1$) would place the energy of GRB 061201 well within the energy distribution of other sGRBs. This physical tension strongly motivates us to re-investigate the field with deeper images.

\subsection{JWST}

\begin{figure*}[t]
    \centering
    \begin{tabular}{cc}
        \includegraphics[width=0.45\textwidth]{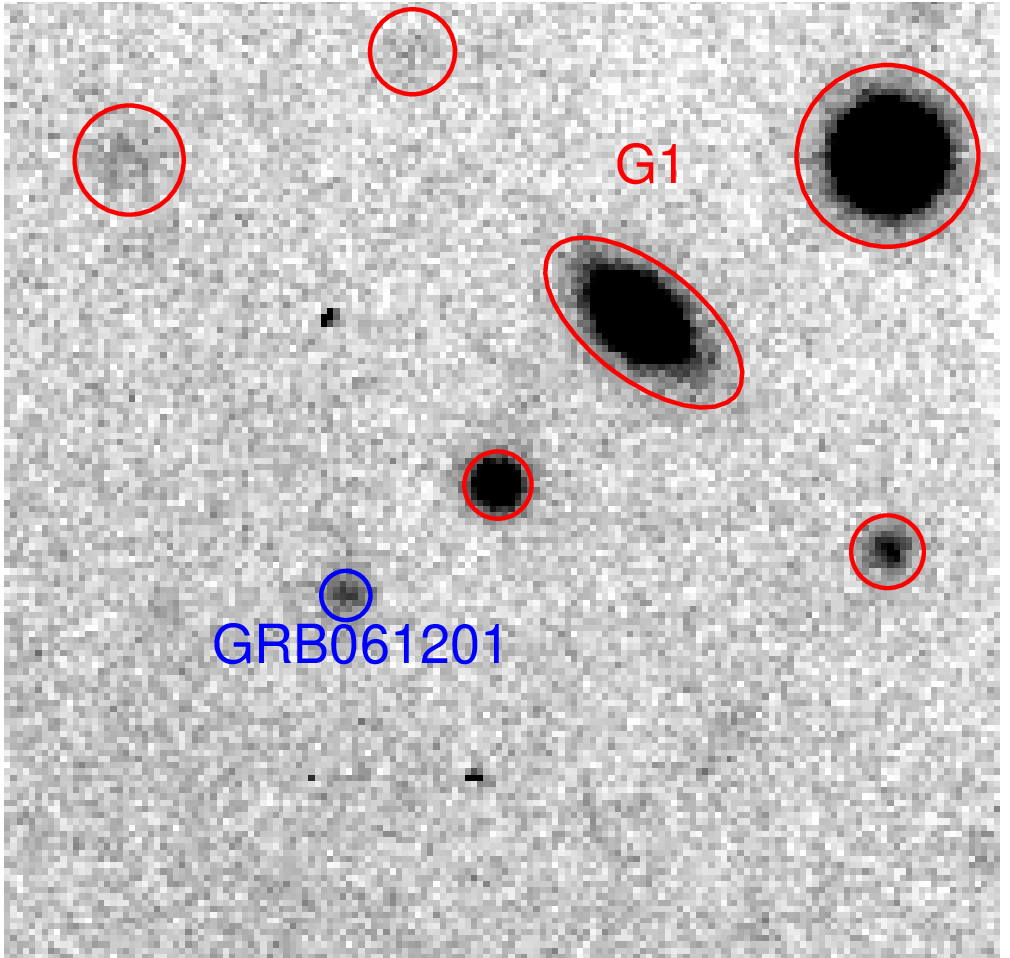} &
        \includegraphics[width=0.45\textwidth]{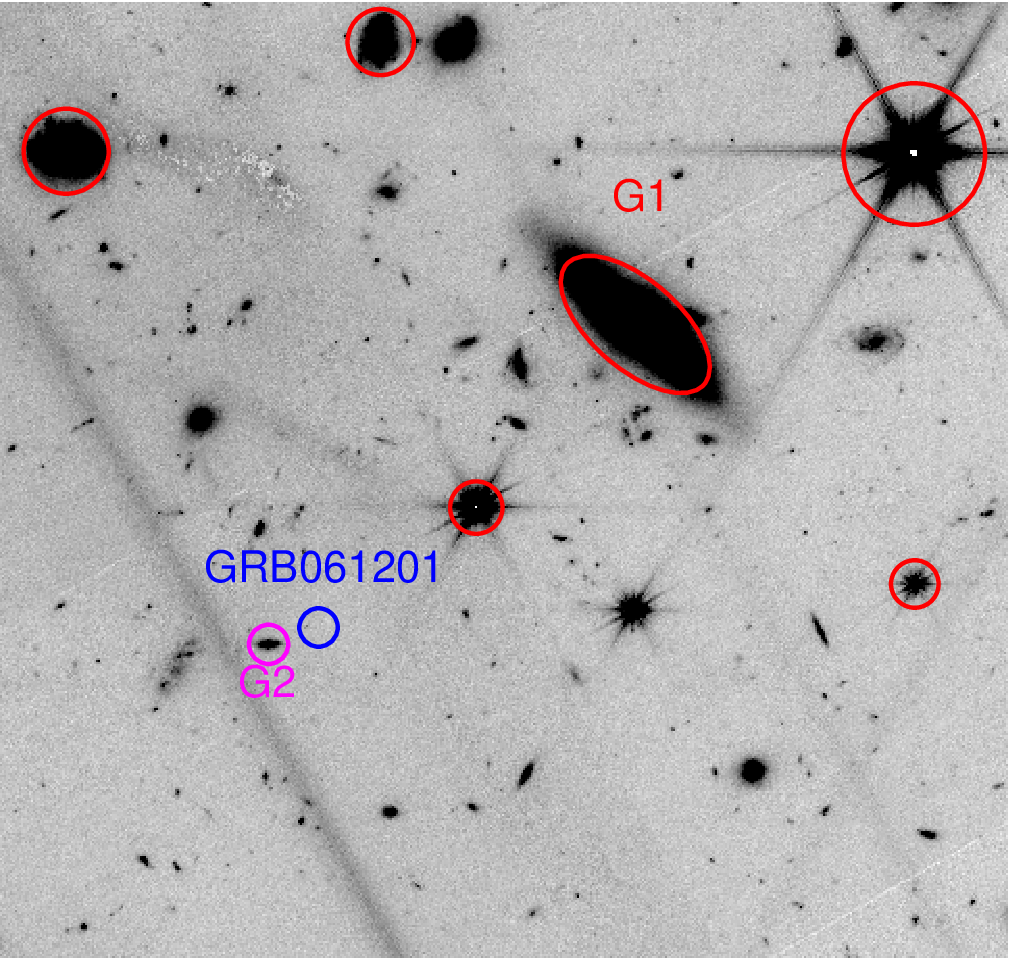} \\
        (a) VLT & (b) JWST \\
    \end{tabular}
    \caption{Field of GRB 061201 observed with VLT and JWST. Red: sources detected in both VLT and JWST images. Blue: position of the optical afterglow. Pink: newly identified host galaxy candidate G2. The images are oriented with North at the top and East to the left.}
    \label{VLT_JWST}
\end{figure*}

We analyzed archival JWST/NIRCam images \citep{2023PASP..135b8001R} covering the field of GRB 061201. Adopting the precise coordinate of the optical afterglow obtained by the VLT as the reference localization, the JWST images reveal a previously unidentified faint galaxy (G2) close to the afterglow (Figure~\ref{VLT_JWST}(b)). Using the pipeline-processed mosaics, we performed aperture-corrected AB photometry in the $F150W2$ and $F322W2$ bands, yielding the magnitudes of $24.73 \pm 0.01 \ {\rm mag}$ and $24.38 \pm 0.01 \ {\rm mag}$, respectively. These magnitudes, combined with the HST photometry described below, are used to constrain the spectral energy distribution (SED) and estimate the photometric redshift (Figures~\ref{JWST_f150w2} and \ref{JWST_f322w2}).

\begin{table*}[htbp]
\centering
\caption{Photometry of the Host Galaxy Candidates G2 and G3}
\label{tab:photometry}

\small
\begin{tabular}{lcccccc}
\hline \hline
Target & F450W & F606W & F814W & F160W & F150W2 & F322W2 \\
 & (mag) & (mag) & (mag) & (mag) & (mag) & (mag) \\
\hline
G2 & $>25.33$ & $26.07 \pm 0.07$ & $25.55 \pm 0.09$ & $24.62 \pm 0.03$ & $24.73 \pm 0.01$ & $24.38 \pm 0.01$ \\
G3 & $>25.22$ & $>26.54$ & $>26.12$ & $>27.11$ & $30.15 \pm 0.21$ & $28.71 \pm 0.15$ \\
\hline
\end{tabular}

\vspace{1ex}

\begin{minipage}{\textwidth}
\small
\textbf{Note.} All magnitudes are in the AB system. The error of F150W2 and F322W2 of G2 is given by the JWST pipeline. The upper limits for all filters are given at the $5\sigma$ level. 
\end{minipage}
\end{table*}

\subsection{HST}

\begin{figure*}[t] 
    \centering
    
    \begin{subfigure}[b]{0.16\textwidth}
        \centering
        \includegraphics[width=\textwidth]{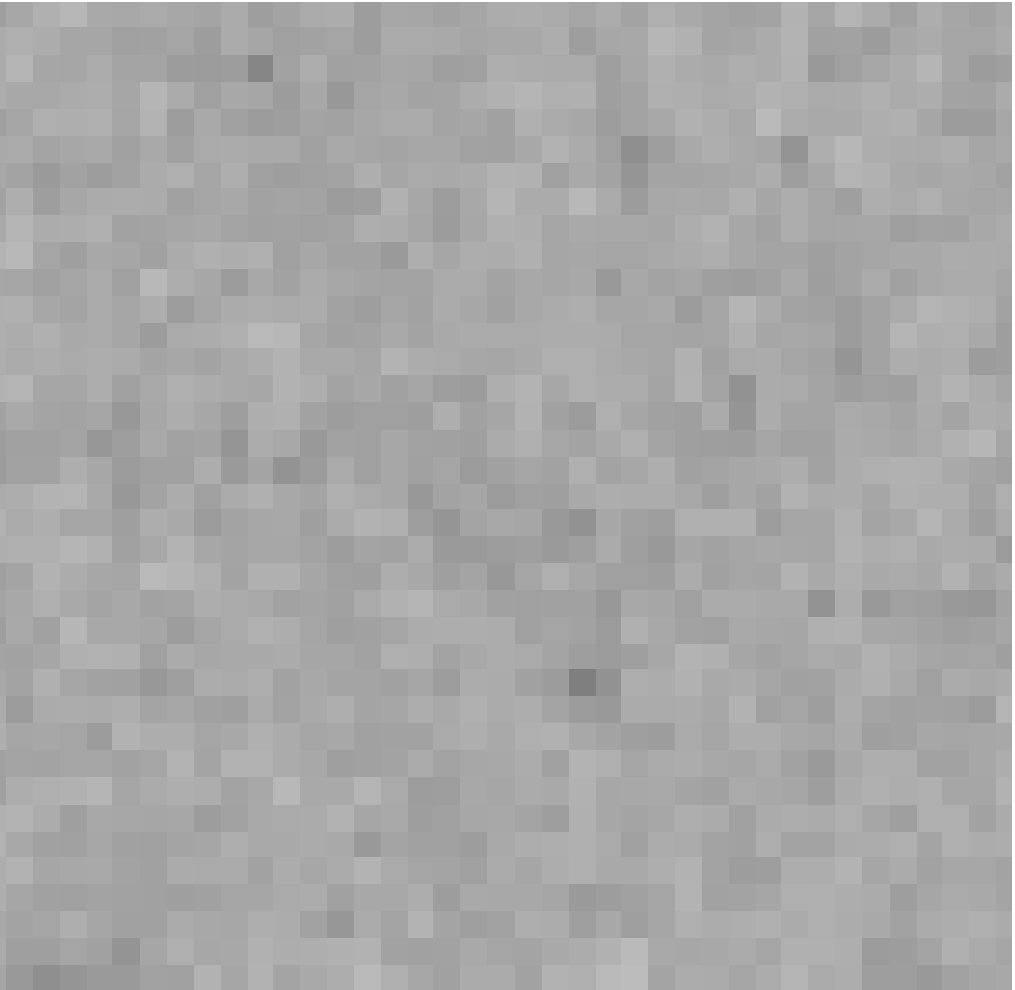}
        \caption{F450W}
        \label{HST_f450w}
    \end{subfigure}
    \hfill
    \begin{subfigure}[b]{0.16\textwidth}
        \centering
        \includegraphics[width=\textwidth]{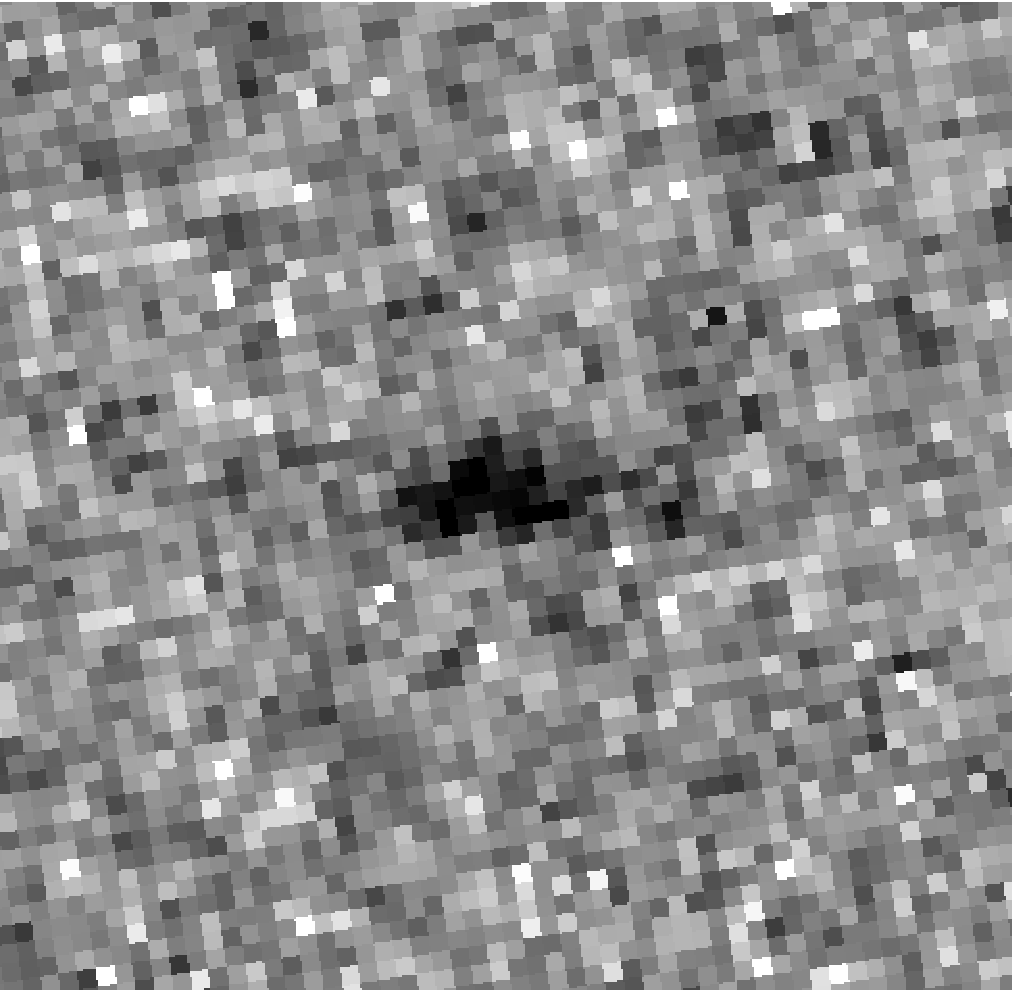}
        \caption{F606W}
        \label{HST_f606w}
    \end{subfigure}
    \hfill
    \begin{subfigure}[b]{0.16\textwidth}
        \centering
        \includegraphics[width=\textwidth]{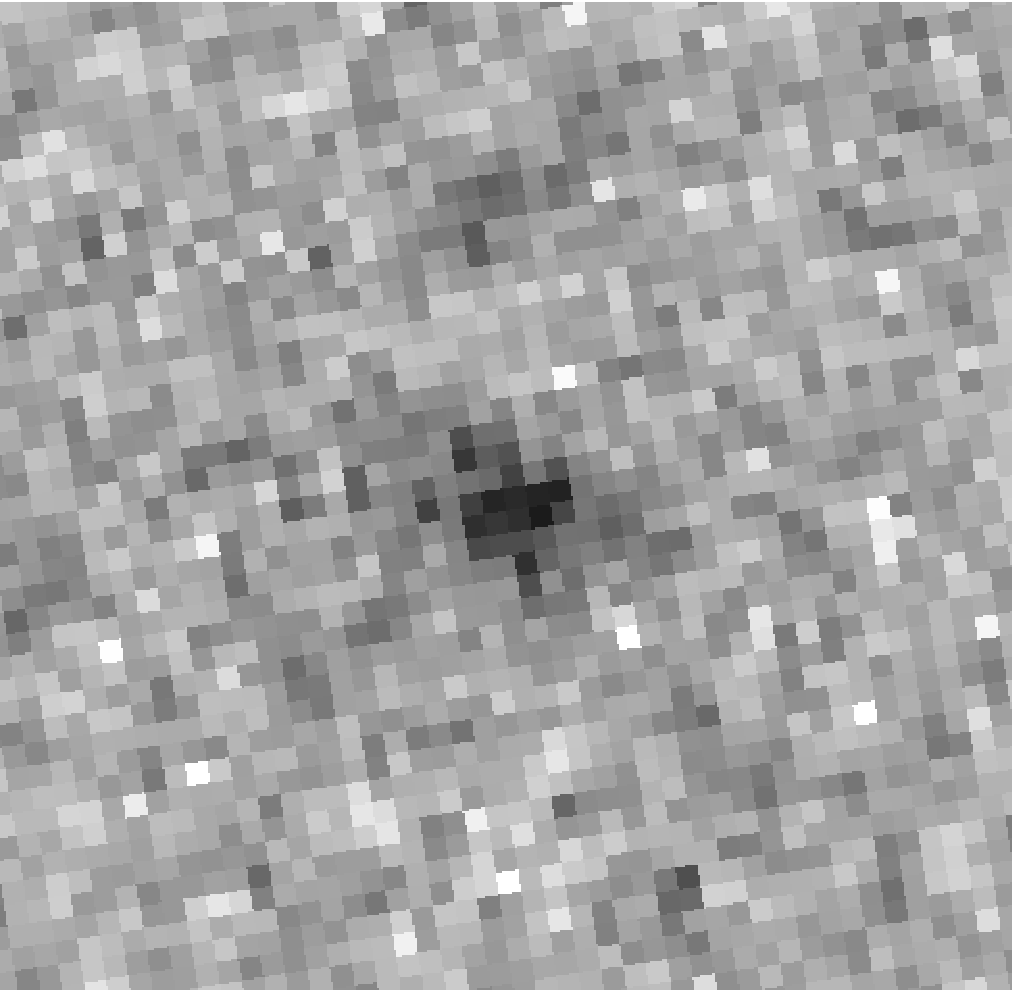}
        \caption{F814W}
        \label{HST_f814w}
    \end{subfigure}
    \hfill
    \begin{subfigure}[b]{0.16\textwidth}
        \centering
        \includegraphics[width=\textwidth]{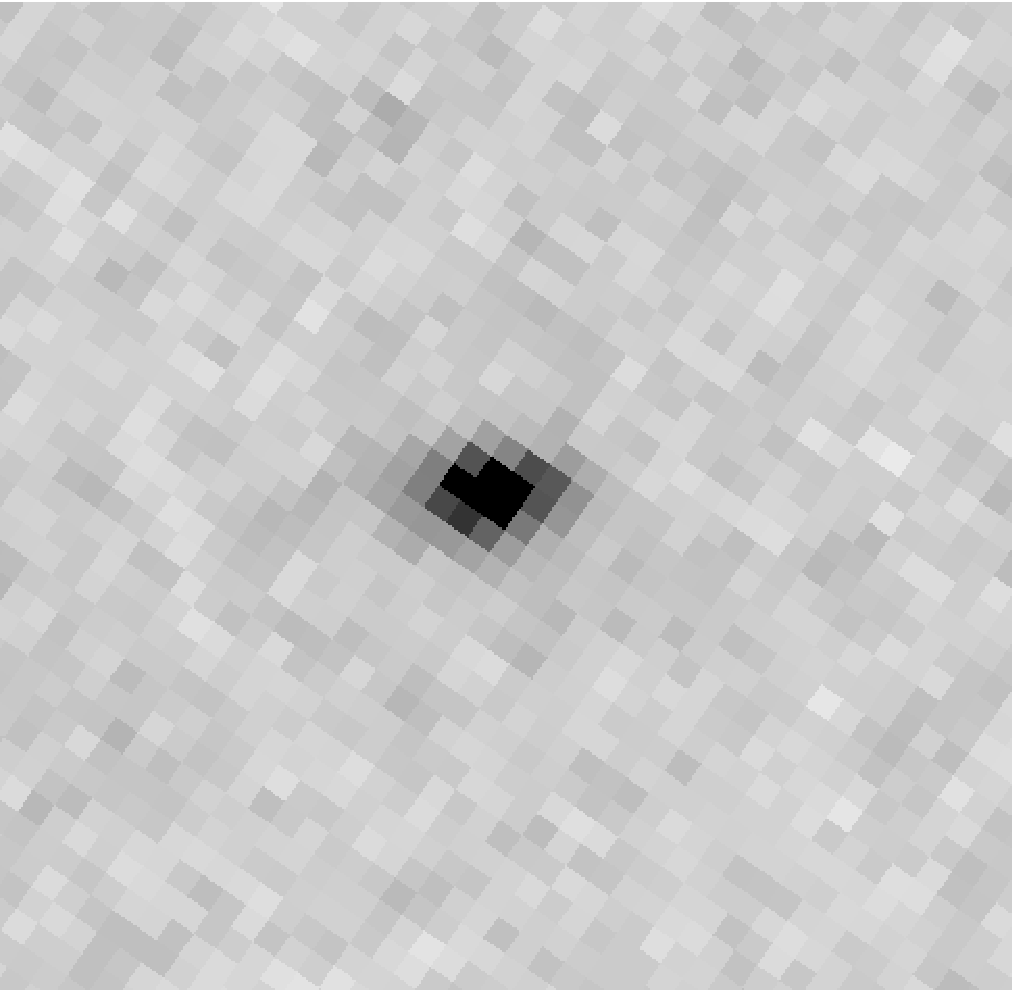}
        \caption{F160W}
        \label{HST_f160w}
    \end{subfigure}
    \hfill
    \begin{subfigure}[b]{0.16\textwidth}
        \centering
        \includegraphics[width=\textwidth]{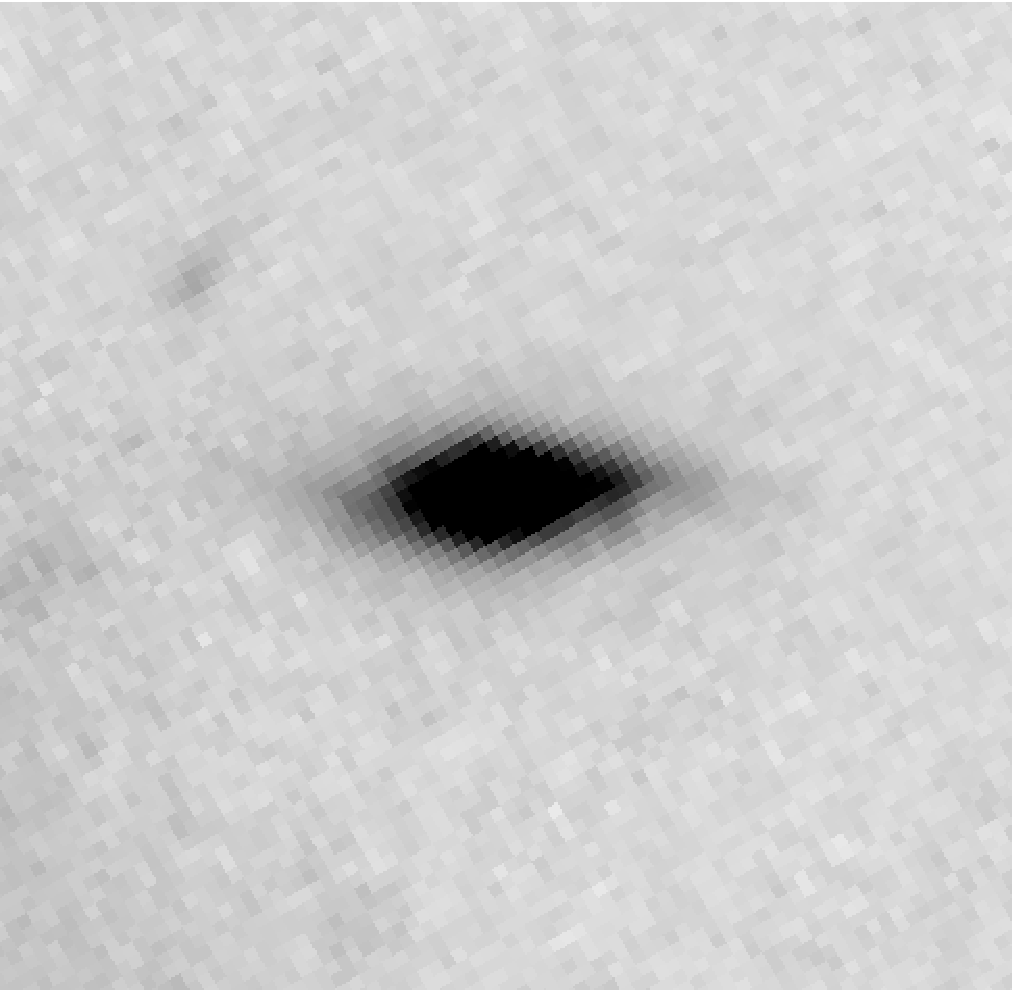}
        \caption{F150W2}
        \label{JWST_f150w2}
    \end{subfigure}
    \hfill
    \begin{subfigure}[b]{0.16\textwidth}
        \centering
        \includegraphics[width=\textwidth]{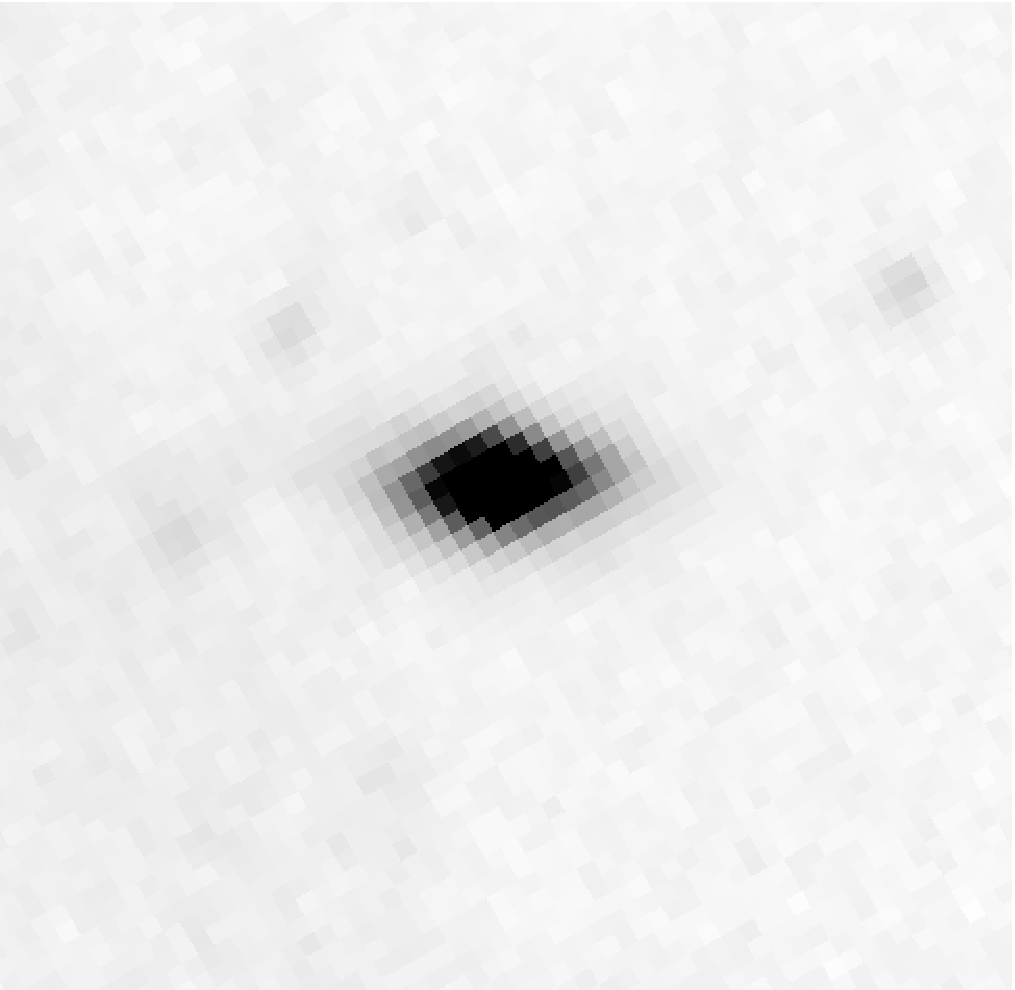}
        \caption{F322W2}
        \label{JWST_f322w2}
    \end{subfigure}
    
    \caption{HST and JWST observations of the host galaxy candidate G2. The images are oriented with North at the top and East to the left.}
    \label{fig:all_bands}
\end{figure*}

We also analyzed archival HST imaging from the WFPC2 \citep{1995PASP..107..156H}, ACS \citep{2005PASP..117.1049S}, and WFC3 \citep{2008SPIE.7010E..1EK} instruments in the $F160W$, $F450W$, $F606W$, and $F814W$ filters. The new candidate host galaxy is robustly detected in the $F160W$ band, marginally detected in $F606W$ and $F814W$ bands, and undetected in the $F450W$ band (Figures~\ref{HST_f160w}, \ref{HST_f450w}, \ref{HST_f606w}, \ref{HST_f814w}). 

Photometry was performed with \texttt{SExtractor} on the background-subtracted science frames. In the high signal-to-noise-ratio (S/N, or SNR) $F160W$ image, the source is spatially resolved, yielding a magnitude of $24.62 \pm 0.03 \ {\rm mag}$. For $F606W$ and $F814W$ bands, we coadded individual exposures (e.g., two consecutive exposures of 1090\,s and 1088\,s in the $F606W$ band). These combined images yield a magnitude of $26.07 \pm 0.07 \ {\rm mag}$ for the $F606W$ band and a magnitude of $25.55 \pm 0.08 \ {\rm mag}$ for the $F814W$ band (which similarly combined two exposures).

As the source is undetected in the $F450W$ band, we derive a $5\sigma$ upper limit with the empirical empty-aperture method. Circular apertures with a radius of $r = 0.50^{\prime\prime}$ are randomly distributed across source-free regions near the target, yielding 60 valid background measurements after masking out nearby objects and image edges. This specific aperture size is carefully chosen to adequately encompass both the instrumental point-spread function (PSF; $\mathrm{FWHM} \sim 0.15^{\prime\prime}$ for WFPC2) and the intrinsic spatial extent of the galaxy ($R_{\rm half} = 0.17^{\prime\prime}$ as resolved in the redder bands). By applying a $3\sigma$ clipped statistic, the standard deviation of these aperture fluxes is adopted as the local background noise. Incorporating the instrumental zero-point ($ZP = 24.04$) and applying the corresponding encircled energy corrections derived from the STScI Instrument Handbooks, we obtain a $5\sigma$ limiting magnitude of $F450W > 25.33 \ {\rm mag}$. The complete multi-band photometry used for our subsequent SED fitting is summarized in Table~\ref{tab:photometry}.

\section{Host Galaxy Candidate}

\subsection{Photometric Redshift Fitting}

\begin{figure*}[t]
    \centering
    \begin{subfigure}[b]{0.6\textwidth}
        \centering
        \includegraphics[width=\textwidth]{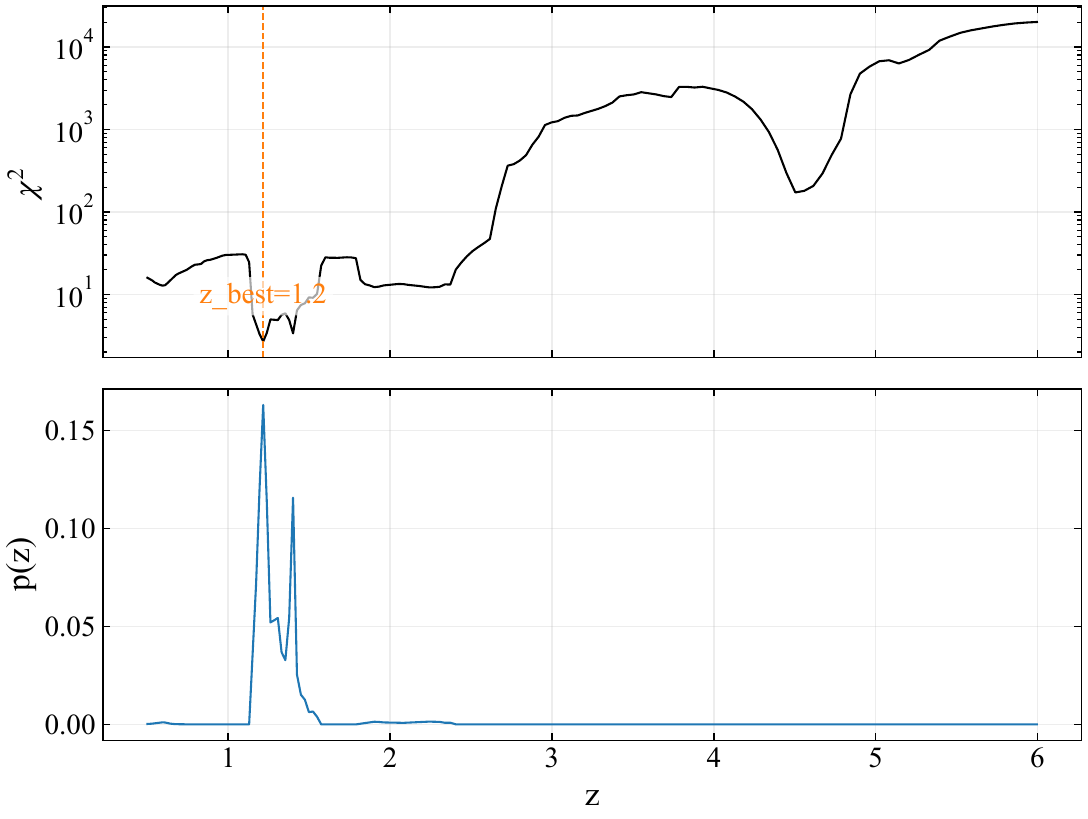}
        \caption{$\chi^2$ and p(z) vs z}
        \label{chi2_pz}
    \end{subfigure}
    
    \begin{subfigure}[b]{0.6\textwidth}
        \centering
        \includegraphics[width=\textwidth]{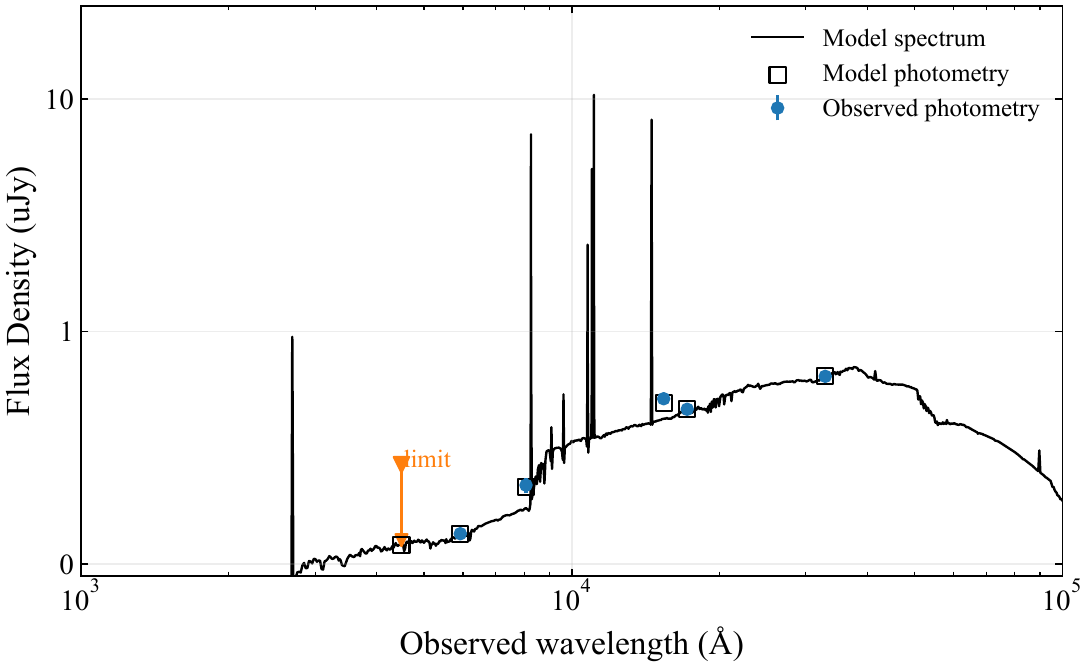}
        \caption{model curve vs observed phot and model phot}
        \label{zphot_model_plot}
    \end{subfigure}
    \caption{Photometric redshift fit for the host galaxy candidate G2}
    \label{z_phot}
\end{figure*}

We utilized the \texttt{EAZY} code to fit the broadband SED constructed from the combined JWST and HST photometry. The resulting $\chi^2(z)$ and probability density, $p(z)$, distributions are shown in Figure~\ref{z_phot}.

The $\chi^2(z)$ distribution exhibits a global minimum at $z\sim1.2$. The marginalized redshift probability density, $p(z)$, is primarily peaked around the global minimum of the $\chi^2(z)$ distribution.

To derive the confidence interval from the asymmetric and multi-peaked $p(z)$, we adopt the highest-posterior-density (HPD) prescription. Applying this method, we obtain a 68\% HPD credible interval of $z \in [1.16, 1.25] \cup [1.28, 1.32] \cup [1.39, 1.41]$ and a 95\% interval of $z \in [1.14, 1.47]$. The best-fit photometric redshift is $z_{\rm best} = 1.2$. This photometric solution is consistent with the hypothesis that the candidate host galaxy lies at a relatively high redshift ($z \gtrsim 1$).

\subsection{Chance-Coincidence Probability}

An observed galaxy in the vicinity of a GRB localization can be either a physically associated host or a chance projection. To quantify the likelihood of a spurious association, we calculate the chance-coincidence probability, $P_{\rm cc}$, following the standard methodology outlined in the literature \citep{2002AJ....123.1111B, 2007ApJ...654..878B, 2008ApJ...677.1157C}. The probability of finding a galaxy of magnitude $m$ or brighter by chance within an effective angular radius $r_{\rm eff}$ is given by:

\begin{equation}
    P_{\rm cc} = 1 - \exp[-\pi r_{\rm eff}^2 \sigma(<m)]
    \label{eq:Pcc}
\end{equation}

where $\sigma(<m)$ is the surface density of galaxies brighter than the candidate's AB magnitude $m$. Following standard convention, the effective search radius $r_{\rm eff}$ depends on the burst offset. If the GRB localization falls within the galaxy's half-light radius, we adopt $r_{\rm eff} = 2R_{\rm half}$. Otherwise, we use $r_{\rm eff} = \max[3\sigma_R, \sqrt{R_0^2 + 4R_{\rm half}^2}]$, where $R_0$ is the angular distance between the GRB position and the center of the galaxy. The total positional uncertainty, $\sigma_R$, is defined as:

\begin{equation}
    \sigma_R = \sqrt{\sigma_{\rm AG}^2 + \sigma_{\rm host}^2}
    \label{eq:sigma_R}
\end{equation}

Here, $\sigma_{\rm AG}$ represents the afterglow localization uncertainty, and $\sigma_{\rm host}$ denotes the uncertainty in the galaxy centroid. To evaluate $\sigma(<m)$, rather than relying on generalized analytic galaxy counts, we empirically estimate the local surface density directly from our imaging. This is calculated by dividing the number of field sources brighter than the candidate, $N(<m)$, by the effective survey area, $A_{\rm eff}$:

\begin{equation}
    \sigma(<m) = \frac{N(<m)}{A_{\rm eff}}
    \label{eq:sigma_m}
\end{equation}

The JWST $F150W2$ image is adopted for the $P_{\rm cc}$ calculation due to its excellent depth and resolution. Adopting the VLT optical afterglow position as the GRB localization with an uncertainty of $0.2^{\prime\prime}$, we measure an angular separation of $R_0 = 1.97^{\prime\prime}$ between the location of the burst and the center of the host candidate. With a measured half-light radius of $R_{\rm half} = 0.17^{\prime\prime}$, the total positional uncertainty is $\sigma_R = 0.30^{\prime\prime}$.

To calculate the local background number counts, we evaluate the $F150W2$ field over an effective unmasked area of $A_{\rm eff} = 3.06\times 10^{4} \ {\rm arcsec}^2$. While restricting counts to extended sources could reduce stellar contamination, it risks excluding compact or high-redshift galaxies, potentially biasing the surface density. We therefore adopt a conservative approach by including all detected objects brighter than the candidate, yielding $N(<m) = 494$ sources. Utilizing Equations (\ref{eq:Pcc}), (\ref{eq:sigma_R}), and (\ref{eq:sigma_m}), we derive a final chance-coincidence probability of $P_{\rm cc} = 0.18$ for this new candidate.

To validate the analytic estimate, we performed a Monte Carlo simulation. We generated $10,000$ random positions across the unmasked image and recorded whether at least one catalog source brighter than the candidate falls within the relevant search radius. The fraction of trials satisfying this criterion yields an empirical chance-coincidence probability of $P_{\rm cc} \approx 0.18$, which is in agreement with the image-based analytic estimate.

While the chance-coincidence probability for G2 ($P_{\rm cc} = 0.18$) exceeds the traditional threshold of $P_{\rm cc} < 0.1$ typically used for secure host identification, this is an expected consequence of deep JWST and HST observations. As imaging depth increases, the surface density of detectable background sources naturally rises, intrinsically yielding larger chance-alignment probabilities for faint targets. A higher threshold (e.g., $P_{\rm cc} < 0.2$) is thus often warranted for such deep fields, as several robust GRB hosts have been securely identified in the literature despite having $P_{\rm cc} > 0.1$ \citep{2022ApJ...940...56F}, making G2 a highly plausible host candidate.

For comparison, the probability for the previously proposed nearby galaxy at $z=0.111$ (G1) is computed. The result shows a formal $P_{\rm cc} = 0.108$ for this galaxy. However, because G1 is relatively bright, its statistical evaluation is limited by small-number statistics; only three sources in the entire field are brighter than it ($N(<m) = 3$). Therefore, this lower $P_{\rm cc}$ value is primarily driven by bright-tail statistics and does not necessarily indicate a more robust physical association.

\subsection{Physical Properties of the Candidate Host}

\begin{figure*}[t]
    \centering
    \begin{tabular}{cc}
        \includegraphics[width=0.45\textwidth]{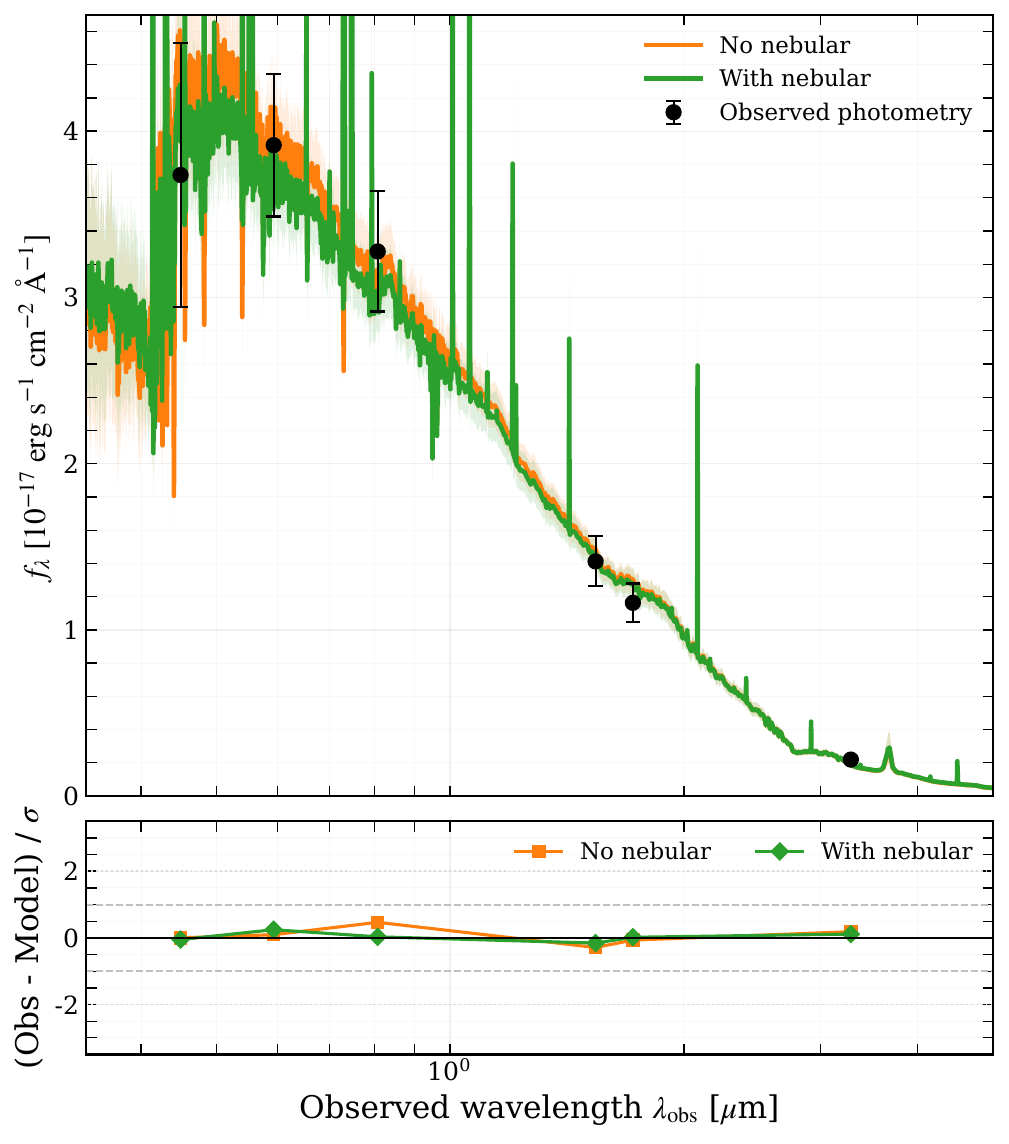} &
        \includegraphics[width=0.45\textwidth]{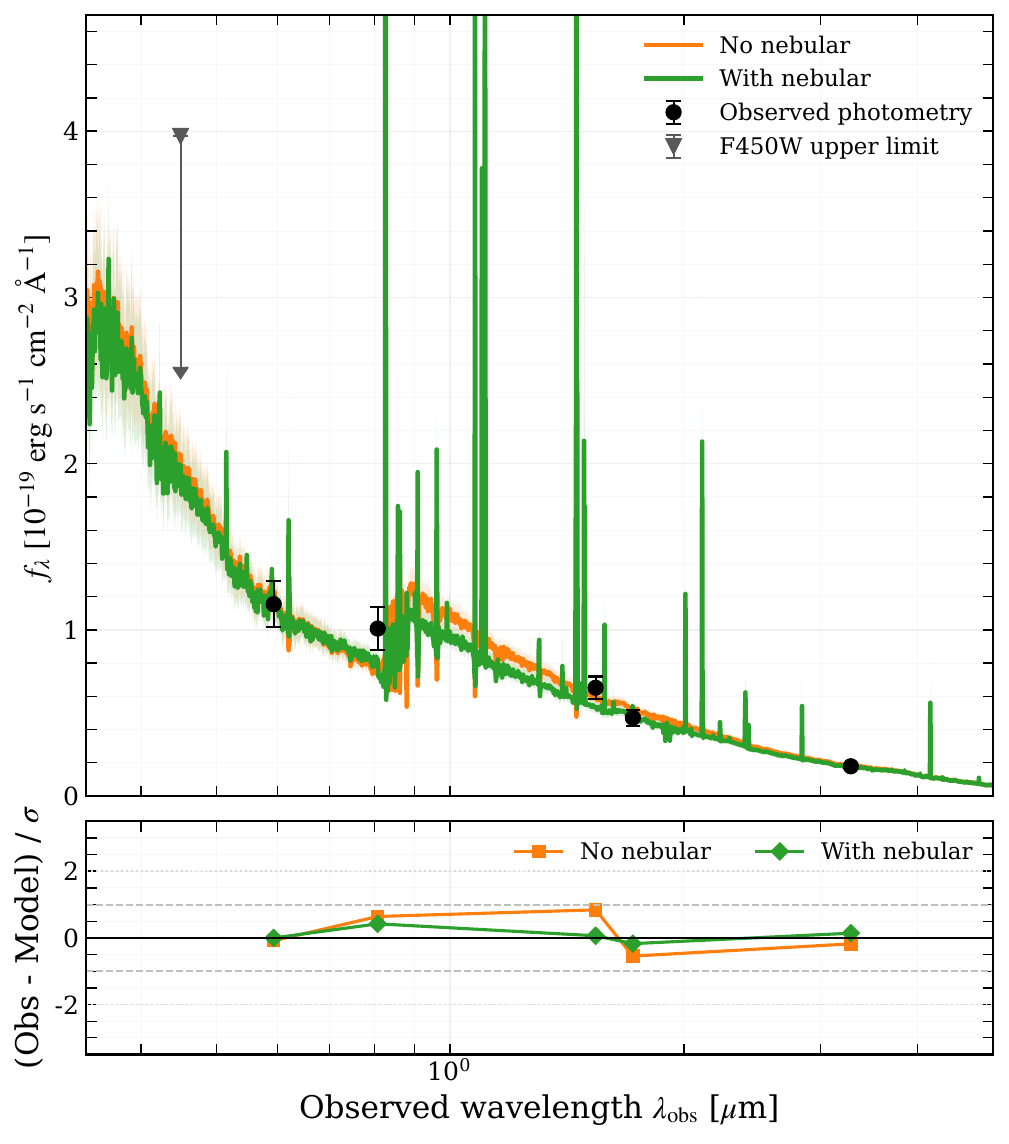} \\
        (a) G1 SED fitting result  & (b) G2 SED fitting result\\
    \end{tabular}
    \caption{SED fitting for G1 and G2 with \texttt{Bagpipes}. The model including nebular emission provides a better fit both for G1 and G2.}
    \label{fig:G2_Bagpipe}
\end{figure*}

In addition to the photometric redshift estimation, we constrained the global physical properties of G2. We employed the \texttt{Bagpipes} code to fit the multi-band SED and derive parameters such as stellar mass and population age. To account for cross-instrument calibration uncertainties, we imposed a 10\% systematic error floor on the photometry and excluded bands with only non-detection upper limits from the fitting procedure. Incorporating nebular emission yields a noticeable improvement in the fit, reducing the total $\chi^2$ to 2.09 compared to the continuum-only model ($\chi^2 = 3.33$) (Figure~\ref{fig:G2_Bagpipe}). This substantial reduction statistically justifies the inclusion of the additional nebular parameter. More importantly, given the active star-forming nature of G2, the presence of nebular emission is physically well-motivated. Furthermore, the blue-end extrapolation of this best-fit model remains consistent with the 5-$\sigma$ upper limit of the $F450W$ band (Figure~\ref{fig:G2_Bagpipe}). Consequently, we adopt the fit with nebular emission as our fiducial model for G2. Because detailed properties of G1 were not provided in prior literature, we applied the same SED fitting procedure to G1 with \texttt{Bagpipes}.

Our best-fit model yields a stellar age of $2.03$\,Gyr for G2 (68\% confidence interval: $0.87$--$3.38$\,Gyr), consistent with a short GRB (sGRB) origin, as the stellar population is sufficiently old to accommodate the extended delay time required for compact binary formation and the subsequent inspiral. Consequently, assuming the progenitor formed during the dominant star-forming episode, we infer a delay time of $t_{\rm delay} \lesssim 2.03$\,Gyr. The SED fitting also yields a stellar mass of $\log(M_*/M_\odot) \approx 8.98$ (68\% confidence interval:$8.87$--$9.09$). The derived star-formation history (SFH) e-folding timescale is $\tau_{\rm SF} = 6.47$\,Gyr, which substantially exceeds the fitted stellar age. This indicates that the galaxy maintains an active star-forming component with sustained optical and ultraviolet emission, facilitating its detection. Finally, we obtain a median visual dust extinction of $A_V = 0.24$\,mag, indicating that G2 is not heavily obscured by dust. For G1, we derive a stellar age of $6.29$\,Gyr (68\% confidence interval: $3.42$--$9.59$\,Gyr), a stellar mass of $\log(M_*/M_\odot) \approx 9.41$ ($9.26$--$9.54$), an SFH e-folding timescale of $\tau_{\rm SF} = 6.83$\,Gyr, and a median visual dust extinction of $A_V = 0.44$\,mag.

\subsection{A Nearby Ultra-faint Galaxy}
\begin{figure*}[t]
  \centering
  \includegraphics[width=0.3\textwidth]{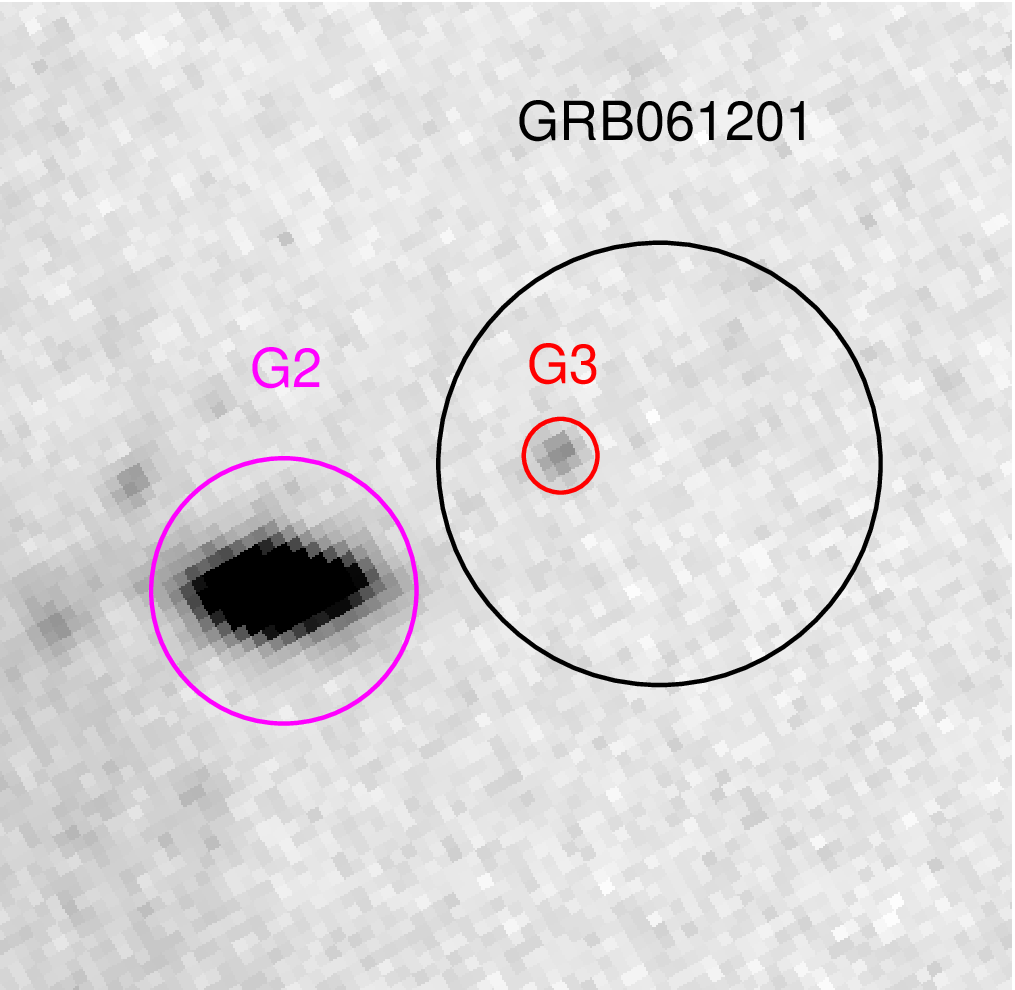}
  \caption{The ultra-faint field source G3 is closer to GRB~061201 in angular distance. The images are oriented with North at the top and East to the left.}
  \label{fig:G3}
\end{figure*}

In addition to our primary host candidate, G2, we identify an ultrafaint source, labeled G3, located at an even smaller projected separation from the GRB 061201 localization (see Figure~\ref{fig:G3}). This object falls below the detection limits of the archival HST observations and is only marginally resolved in the deep JWST images.

Centered at coordinates RA = 22:08:32.211, Dec = -74:34:46.04, G3 is situated at an angular separation of $0.49^{\prime\prime}$ from the GRB position. We evaluated the chance-coincidence probability for G3 using the $F322W2$ image, where the source is detected with the highest signal-to-noise ratio. Extracting the local background source density ($N(<m) = 4853$ across an effective area $A_{\rm eff} = 2.99\times10^{4} \ {\rm arcsec}^2$) and adopting an effective radius $R_{\rm half} = 0.10^{\prime\prime}$ with a positional uncertainty $\sigma_R = 0.35^{\prime\prime}$, we derive a probability of $P_{\rm cc} = 0.43$. This value is substantially higher than that of G2, statistically disfavoring G3 as the true host galaxy.

\begin{figure*}[t]
    \centering
    
    \begin{subfigure}[b]{0.16\textwidth}
        \centering
        \includegraphics[width=\textwidth]{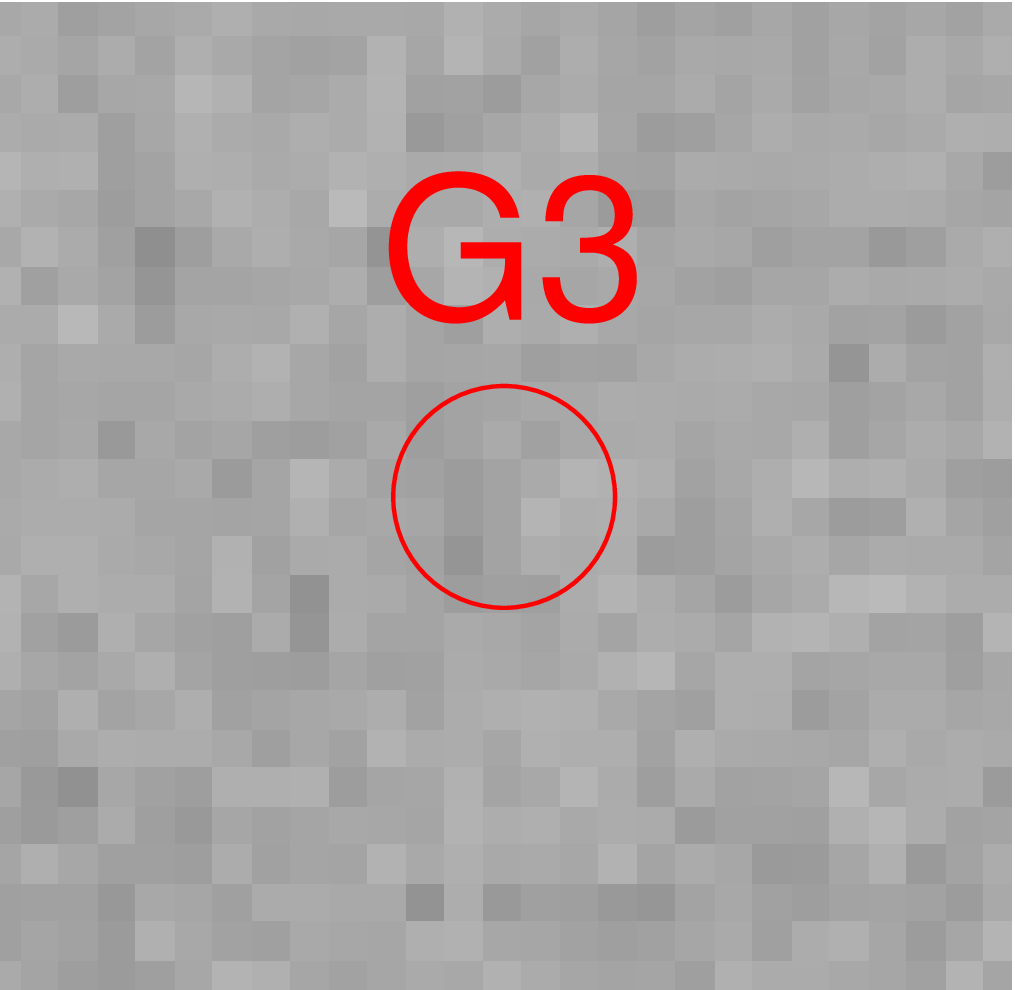}
        \caption{F450W}
        \label{HST_f450w_G3}
    \end{subfigure}
    \hfill
    \begin{subfigure}[b]{0.16\textwidth}
        \centering
        \includegraphics[width=\textwidth]{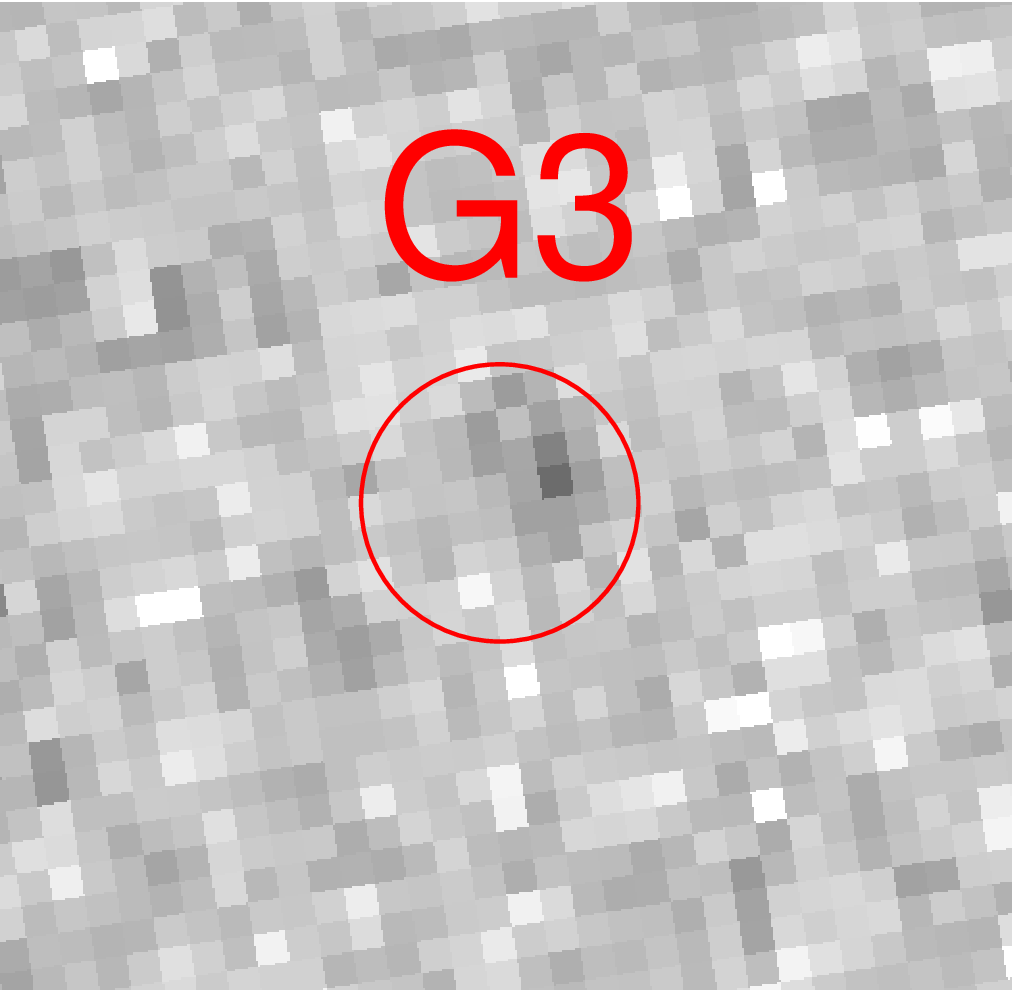}
        \caption{F606W}
        \label{HST_f606w_G3}
    \end{subfigure}
    \hfill
    \begin{subfigure}[b]{0.16\textwidth}
        \centering
        \includegraphics[width=\textwidth]{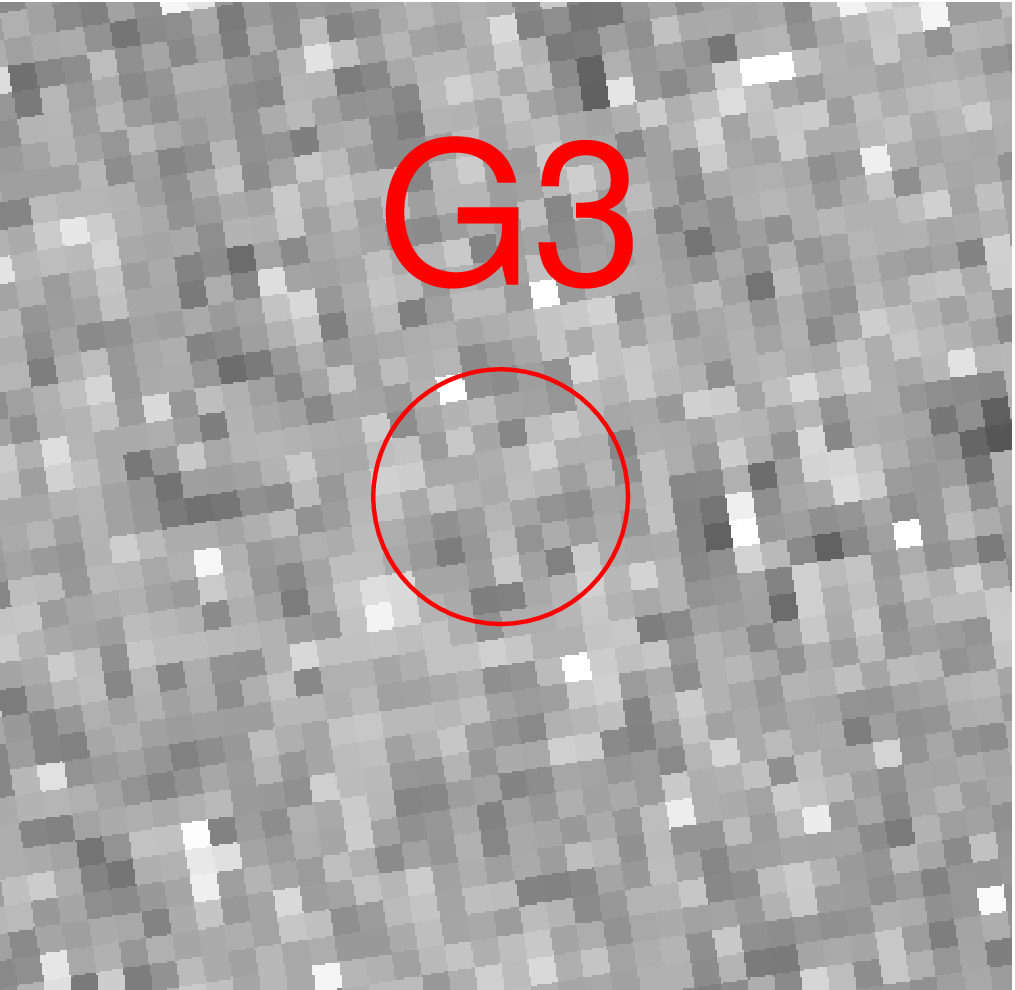}
        \caption{F814W}
        \label{HST_f814w_G3}
    \end{subfigure}
    \hfill
    \begin{subfigure}[b]{0.16\textwidth}
        \centering
        \includegraphics[width=\textwidth]{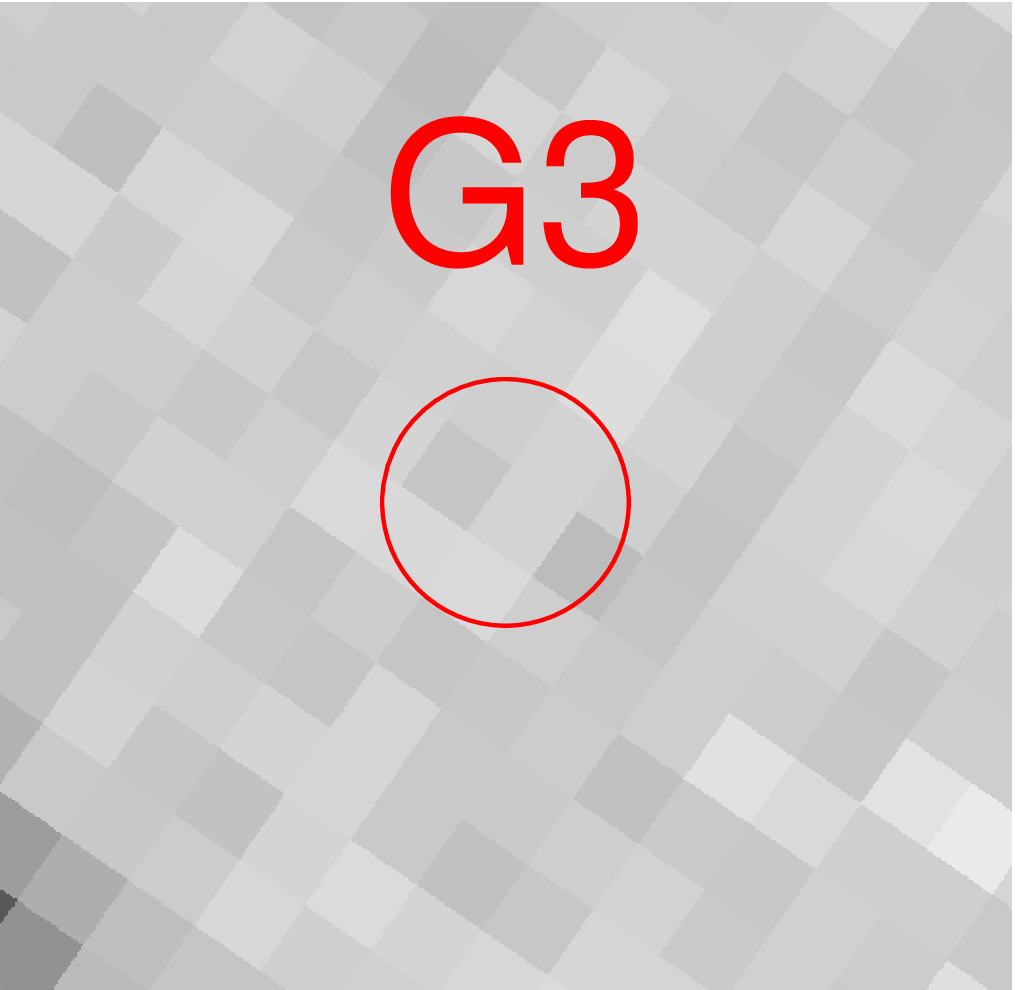}
        \caption{F160W}
        \label{HST_f160w_G3}
    \end{subfigure}
    \hfill
    \begin{subfigure}[b]{0.16\textwidth}
        \centering
        \includegraphics[width=\textwidth]{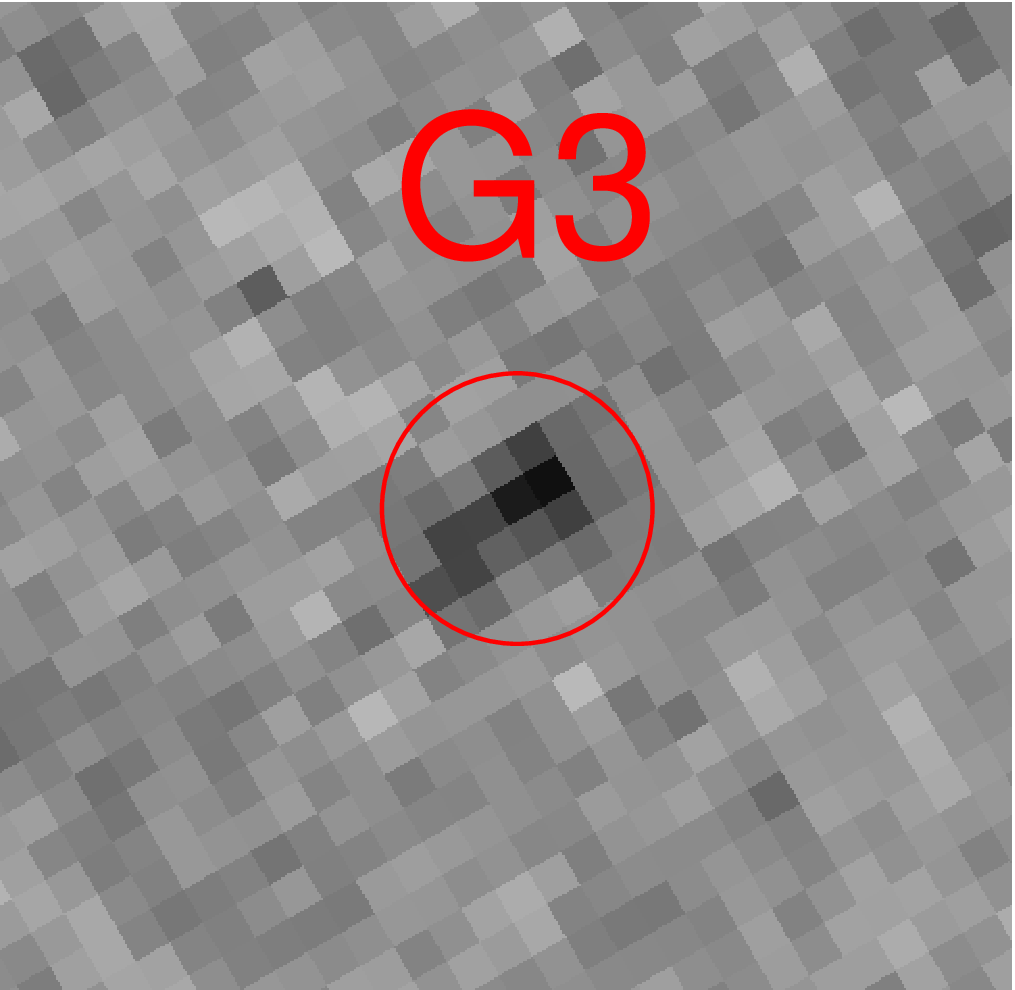}
        \caption{F150W2}
        \label{JWST_f150w2_G3}
    \end{subfigure}
    \hfill
    \begin{subfigure}[b]{0.16\textwidth}
        \centering
        \includegraphics[width=\textwidth]{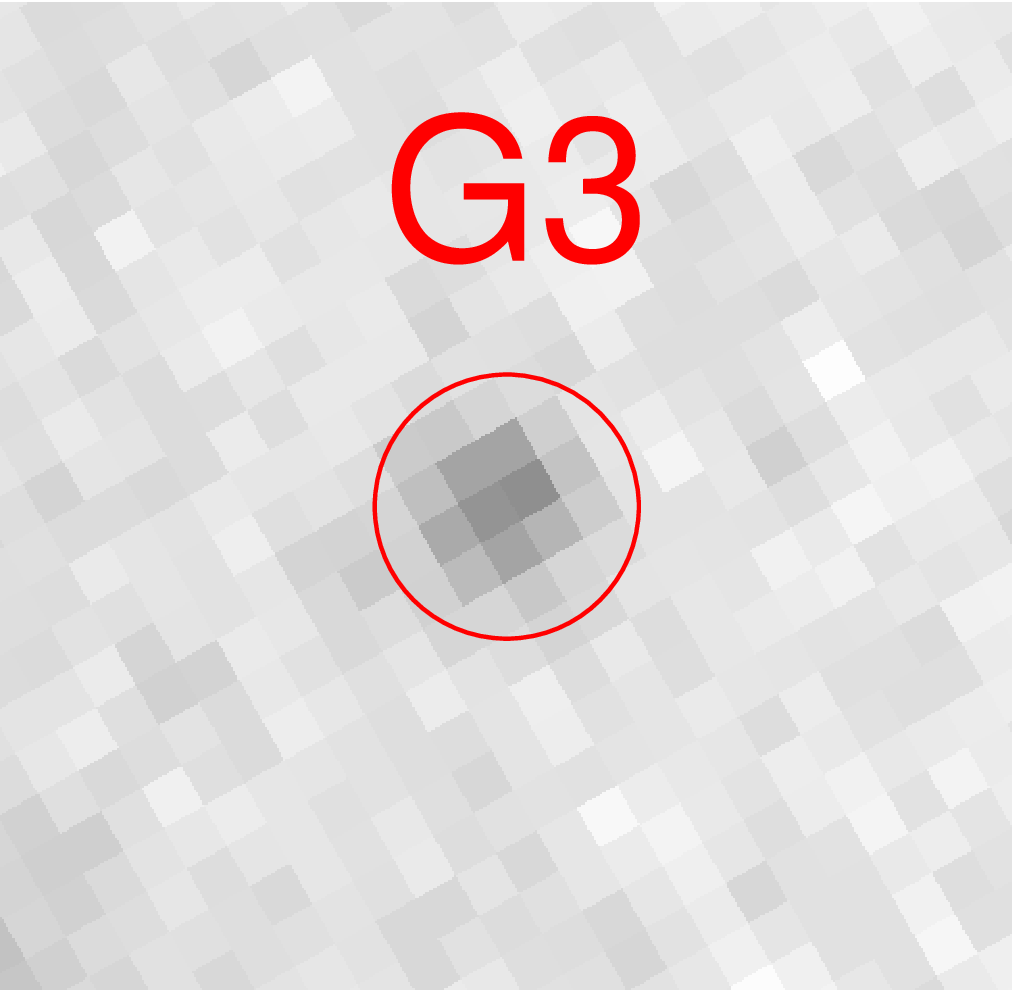}
        \caption{F322W2}
        \label{JWST_f322w2_G3}
    \end{subfigure}
    
    \caption{Images of G3 obtained by HST and JWST. The images are oriented with North at the top and East to the left.}
    \label{fig:G3_in_JWSTandHST}
\end{figure*}

Our photometric analysis of this faint galaxy yields detections of $F150W2 = 30.15 \pm 0.21 \ {\rm mag}$ and $F322W2 = 28.71 \pm 0.15 \ {\rm mag}$. For the filters in which G3 remained undetected, $5\sigma$ upper limits were derived following the same empirical empty-aperture method used for G2. We adopted circular apertures with a radius of $r = 0.20^{\prime\prime}$, which securely encompasses both the galaxy's compact morphology ($R_{\rm half} = 0.10^{\prime\prime}$) and the broadest instrumental PSF ($\mathrm{FWHM} \lesssim 0.15^{\prime\prime}$) across these filters. Consequently, the detection limits were determined as $F450W > 25.22$, $F606W > 26.54$, $F814W > 26.12$, and $F160W > 27.11 \ {\rm mag}$ (see Figure~\ref{fig:G3_in_JWSTandHST} and Table~\ref{tab:photometry}). Due to the limited signal-to-noise ratio, a precise photometric redshift for G3 cannot be uniquely determined; however, the data reveal a significant color break between the $F150W2$ and $F322W2$ bands. Whether this color break is attributed to the Lyman break, the Balmer break, or the $4000$~\AA\ break, it consistently indicates a high redshift of $z > 3$. Conversely, optical spectroscopy of the GRB 061201 afterglow, despite the low signal-to-noise ratio, shows no evidence of a Ly$\alpha$ break at wavelengths $\lambda > 4000$~\AA \citep{2007A&A...474..827S}, placing a robust upper limit on the burst's redshift at $z < 2.3$.

Given this apparent discrepancy in their respective redshift ranges, we can strongly disfavor G3 as the host galaxy of GRB 061201. Despite the unprecedented sensitivity and spatial resolution of JWST, we detect no alternative or closer host candidates within the localization region. Next, we will discuss the likelihood of G1 and G2 serving as the host galaxy.

\section{Comparison of the candidates}
\subsection{Physical properties and offset}

\begin{figure*}[t]
  \centering
  \includegraphics[width=0.6\textwidth]{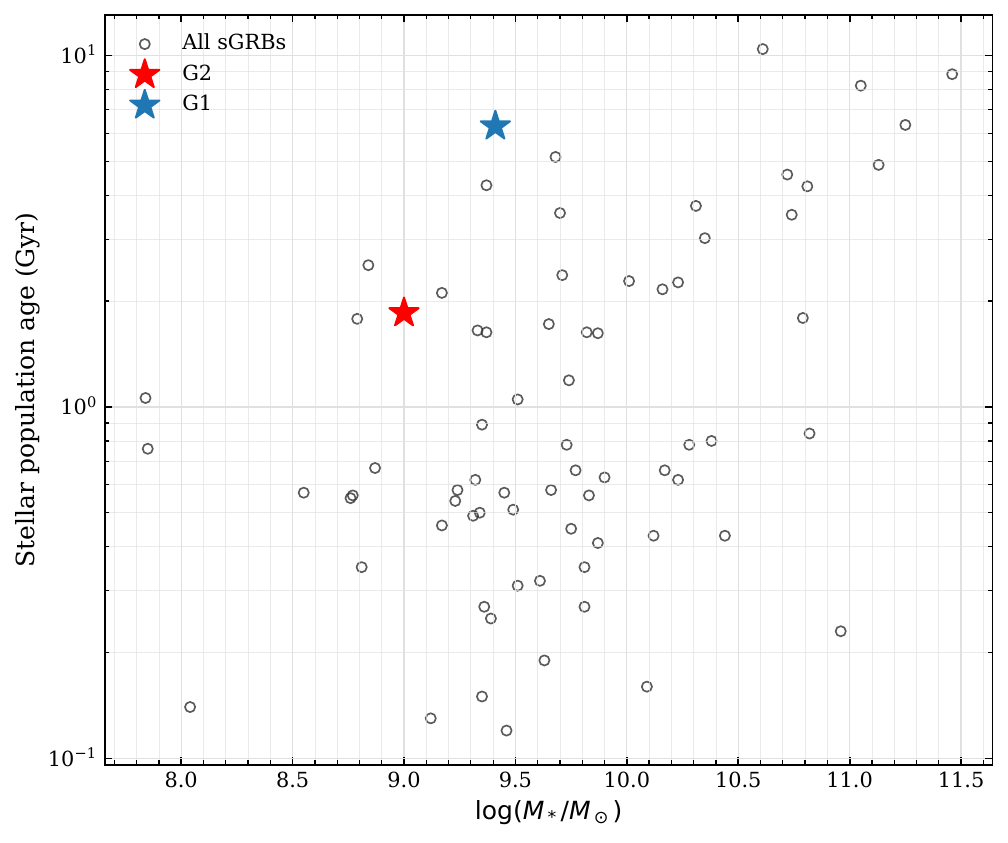}
  \caption{Stellar mass and age of G1 (blue star) and G2 (red star) in the context of sGRB host galaxies (gray circles).}
  \label{fig:mass_age}
\end{figure*}

Our SED modeling yields a stellar mass of $\log(M_*/M_\odot) \approx 8.98$ for G2. Statistically, a substantial fraction of sGRBs are found in similarly low-mass galaxies, making G2 a representative host environment. Importantly, this low mass implies a relatively shallow gravitational potential well; thus, a compact binary receiving an asymmetric natal kick during supernova formation is more likely to exceed the local escape velocity and travel to the outskirt of the galaxy. This dynamical scenario provides a plausible mechanism for the substantial physical offset observed for this event. 
To further contextualize our findings, we compare the stellar mass and age of G2 (and G1) against the established sGRB host galaxy population \citep{2022ApJ...940...56F,2022ApJ...940...57N}, as illustrated in Figure~\ref{fig:mass_age}. This comparison indicates that the macroscopic properties of G2 are more consistent with the core demographic distribution of known sGRB hosts than G1.

\begin{figure*}[t]
    \centering
    \begin{subfigure}[b]{0.75\textwidth}
        \centering
        \includegraphics[width=\textwidth]{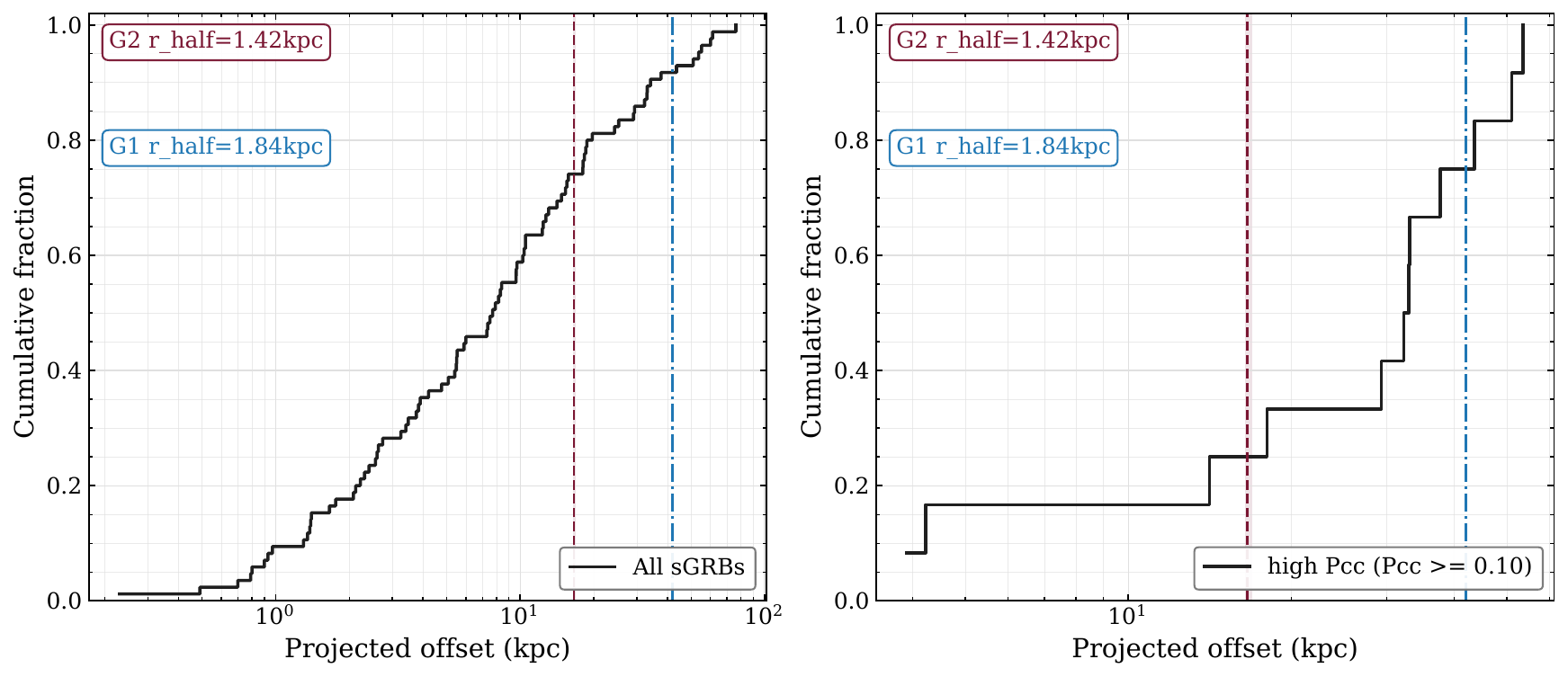}
        \caption{offset comparison}
        \label{fig:offset}
    \end{subfigure}
    
    \begin{subfigure}[b]{0.4\textwidth}
        \centering
        \includegraphics[width=\textwidth]{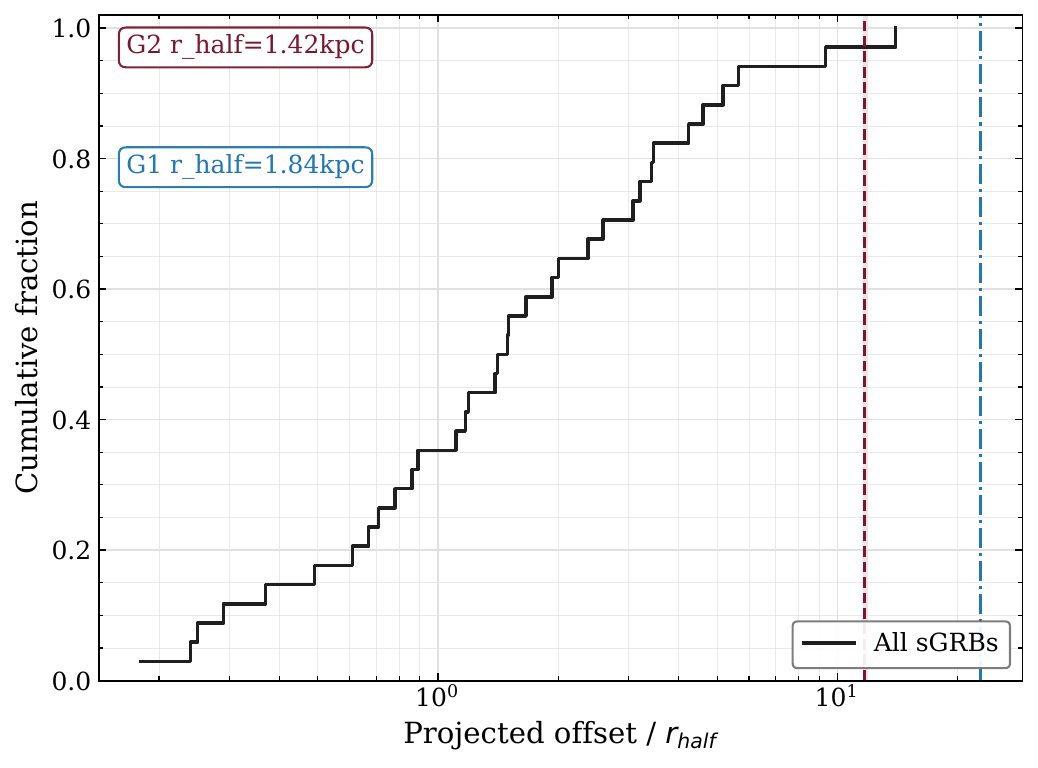}
        \caption{normalized offset comparison}
        \label{fig:normoffset}
    \end{subfigure}
    \caption{(a) Physical offset distributions for all sGRBs (left panel) and the sub-sample with high $P_{\rm cc}$ values (right panel). (b)Comparison of the normalized offsets.}
    \label{offset_results}
\end{figure*}

We also compared the projected physical offset of GRB 061201 with the sGRB population adopted from \cite{2022ApJ...940...56F}. The physical offset is $16.42$--$16.91$\,kpc at the 95\% confidence level for G2. Combined with the inferred progenitor delay time of $0.1$--$2.03$\,Gyr, the required natal kick velocity is approximately $10$--$205 \ {\rm km \ s^{-1}}$, falling well within theoretical expectations for sGRB progenitors. 

In addition to the physical offset, we evaluated the host-normalized offset (defined as the physical offset divided by the effective half-light radius, $R_{\rm half}$; Figure~\ref{fig:offset} left), which provides a more robust metric for the burst's relative position within its host. Given that G2 and G1 have a chance-coincidence probability of $P_{\rm cc} > 0.1$, we constructed a control subsample of known sGRBs whose host galaxies exhibit similarly high $P_{\rm cc}$ values (\ref{fig:offset} right). Because reliable $R_{\rm half}$ measurements are scarce for this specific subset (with only two such events available), we omitted the normalized offset comparison for the high-$P_{\rm cc}$ sample to avoid small-number statistical biases (\ref{fig:normoffset}. 

The results demonstrate that, assuming G2 is the actual host, the physical offset of GRB 061201 falls within the distributions of both the entire sGRB population and the specific $P_{\rm cc} > 0.1$ subset. While the normalized offset of GRB 061201 resides at the higher end of the broader population, it remains well within the observed distribution and does not constitute a statistically significant outlier. In contrast, the offset of G1 resides in the outer tail of the sGRB population distribution. Notably, in the normalized offset distribution, G1 emerges as a significant outlier, lying beyond the range of the core population. Therefore, we conclude that G2 presents a more self-consistent and plausible host environment for GRB 061201 than G1.

\subsection{Burst Energetics and Afterglow Modeling}
\begin{figure*}[t]
    \centering
    \begin{tabular}{cccc}
        \includegraphics[width=0.45\textwidth]{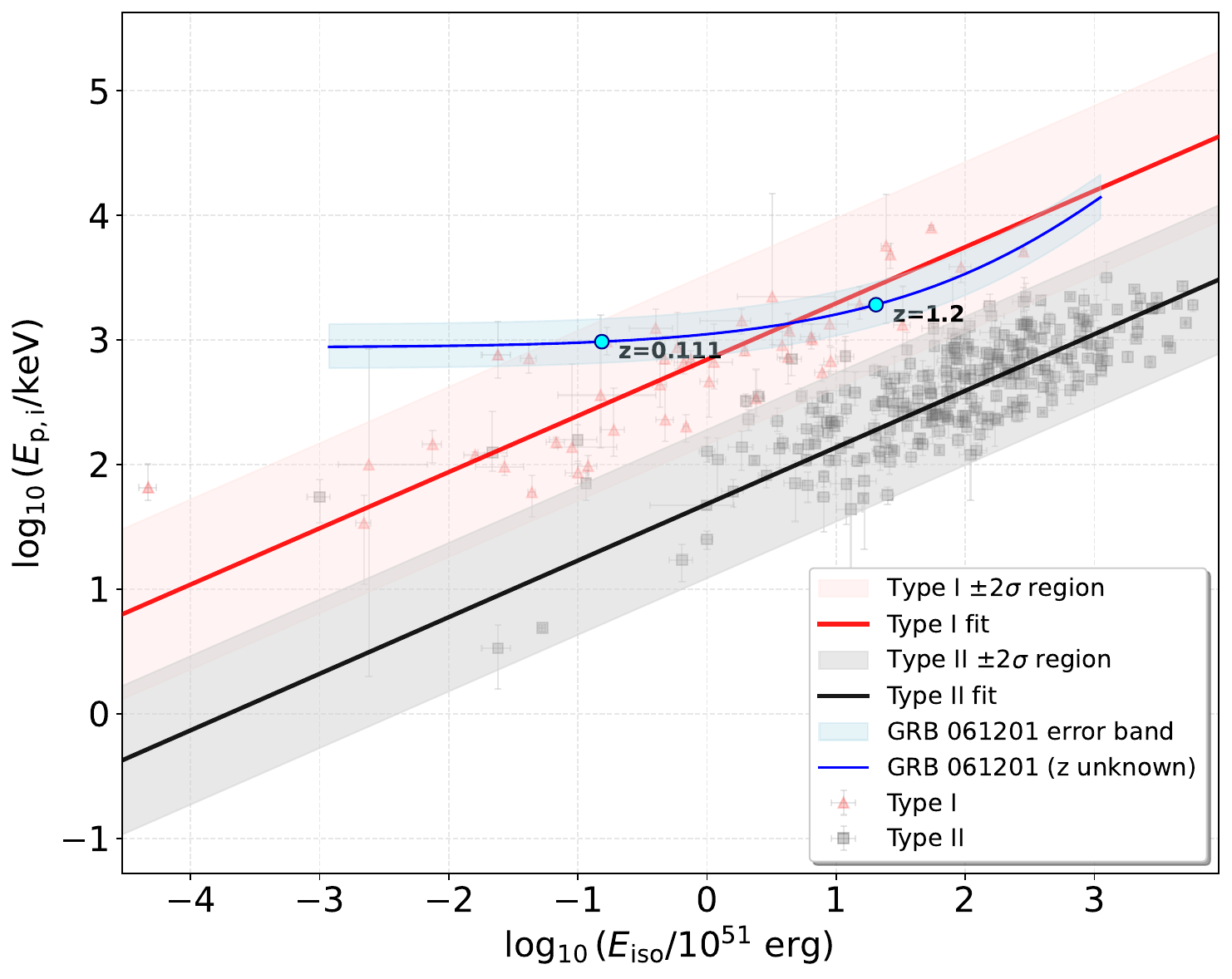} &
        \includegraphics[width=0.45\textwidth]{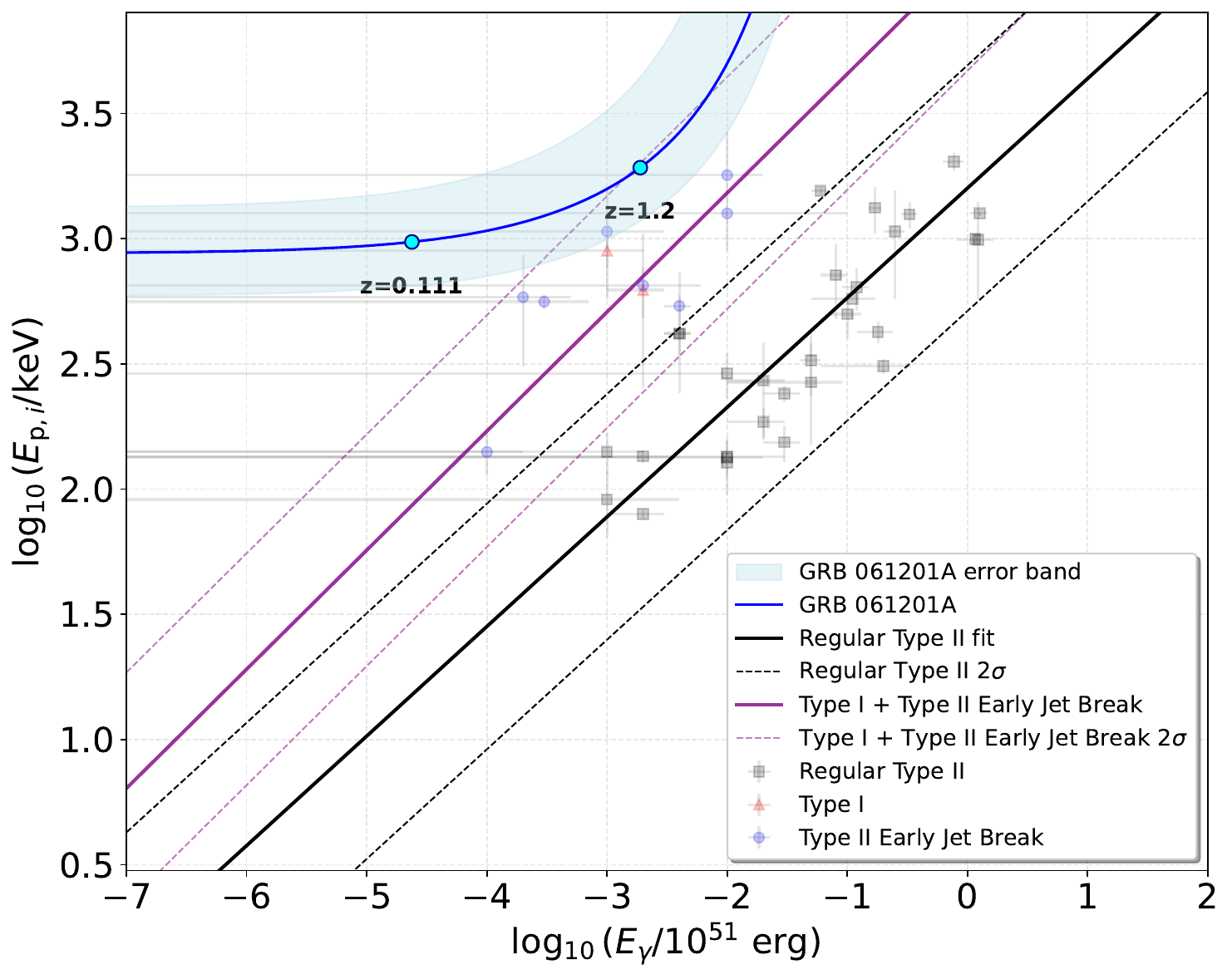} \\
        (a) $E_{p,i}-E_{iso}$ & (b) $E_{p,i}-E_\gamma$ \\
    \end{tabular}
    \caption{Energy correlations of GRB 061201 with different redshift. Left panel (a):$E_{p,i}$ vs $E_{iso}$. Right panel (b): $E_{p,i}$ vs $E_\gamma$ which is beaming-corrected.}
    \label{fig:Amati and Ghirlanda}
\end{figure*}

To address the potential energetic tension discussed earlier, we evaluated the position of GRB 061201 on the $E_{\rm p,i}$--$E_{\rm iso}$ \citep{2002A&A...390...81A} and $E_{\rm p,i}$--$E_\gamma$ \citep{2004ApJ...616..331G} correlation planes under the two redshift assumptions. For the $E_{\rm p,i}$--$E_{\rm iso}$ relation, both the $z=0.111$ and $z=1.2$ scenarios place the burst well within the established distribution of sGRBs (Figure~\ref{fig:Amati and Ghirlanda}(a)). However, once the beaming correction is applied under the jet-break interpretation, only the $z=1.2$ solution remains consistent with the $E_{\rm p,i}$--$E_\gamma$ relation for short bursts within the $2\sigma$ confidence region. Under the $z=0.111$ assumption, the beaming-corrected energy, $E_\gamma$, falls significantly below the expected value, placing the burst well outside the standard sGRB region at more than the $2\sigma$ level
(Figure~\ref{fig:Amati and Ghirlanda}(b)). Consequently, burst energetics strongly suggest that the local-origin hypothesis is posing a substantial challenge to the $z=0.111$ association.

To provide a physical diagnostic for the redshift of GRB 061201, we performed a global joint fit to its broad-band afterglow data. In preparation for this multi-band modeling, we re-processed the complete \textit{Swift}/UVOT dataset, incorporating all available observations from \cite{2007A&A...474..827S}, and systematically rebinned the photometry to avoid unduly broad time spans within individual bins. Additionally, the archival SOAR $J$-band and $K$-band measurements were compiled and presented in Table~\ref{tab:observation}. We modeled the afterglow emission with the code \texttt{VEGASAFTERGLOW} and the Markov Chain Monte Carlo (MCMC) algorithm to evaluate both the $z=0.111$ and $z=1.2$ scenarios.

\begin{table*}[htbp]
\centering
\caption{Photometry for Swift UVOT and SOAR observations of GRB 061201}
\label{tab:observation}
\begin{tabular}{lccc}
\hline \hline
Filter	& T$_{\rm start}$ & T$_{\rm elapse}$	& Mag(Limit) \\
	& (second)	 & (second)	& (Vega) \\
\hline
\hline
uvv	& 68.899	& 9.3926	& $>17.51$ \\
uwh	& 88.129	& 99.770	& $21.36\pm0.48$ \\
uvv	& 193.32	& 399.77	& $>20.16$ \\
um2	& 598.89	& 19.775	& $>17.51$ \\
uw1	& 623.12	& 19.763	& $>18.08$ \\
uuu	& 646.91	& 19.763	& $>18.69$ \\
ubb	& 671.16	& 9.7560	& $>18.61$ \\
uwh	& 684.83	& 9.7678	& $>19.50$ \\
uw2	& 699.55	& 19.764	& $>18.00$ \\
uvv	& 723.30	& 10.859	& $>17.66$ \\
um2	& 5385.7	& 199.77	& $20.93\pm0.76$ \\
uw1	& 5590.5	& 199.78	& $20.45\pm0.46$ \\
uuu	& 5794.9	& 199.77	& $20.83\pm0.52$ \\
ubb	& 5999.7	& 199.77	& $>20.72$ \\
uwh	& 6204.0	& 199.77	& $21.22\pm0.30$ \\
uw2	& 6409.4	& 104.42	& $19.84\pm0.41$ \\
uvv	& 9833.9	& 2304.1	& $>20.31$ \\
um2	& 12143 	& 3806.9	& $>20.42$ \\
uw1 & 17619 	& 308.24 	& $>20.45$ \\
uw1	& 21394 	& 37049	 	& $22.45\pm0.35$ \\
J 	& 41212.8 	& 4200 		& $21.51\pm0.13$ \\
K 	& 49420.8 	& 2160 		& $>19.8$ \\
uw1	& 86429 	& 74044 	& $>22.82$ \\
J 	& 125452.8 	& 5400 		& $>22.9$ \\
\hline
\end{tabular}

\vspace{0.2cm}
\begin{minipage}{0.65\textwidth}
\raggedright \small
\textbf{Note:}\\
$a$. Magnitudes are not corrected for the Galactic extinctions of $E(B-V)=0.0647$ (\citep{2011ApJ...737..103S}) at this position.\\
$b$. The limiting magnitude is set at $3\sigma$.
\end{minipage}
\end{table*}

\begin{figure*}[t]
    \centering
    \includegraphics[width=0.48\linewidth]{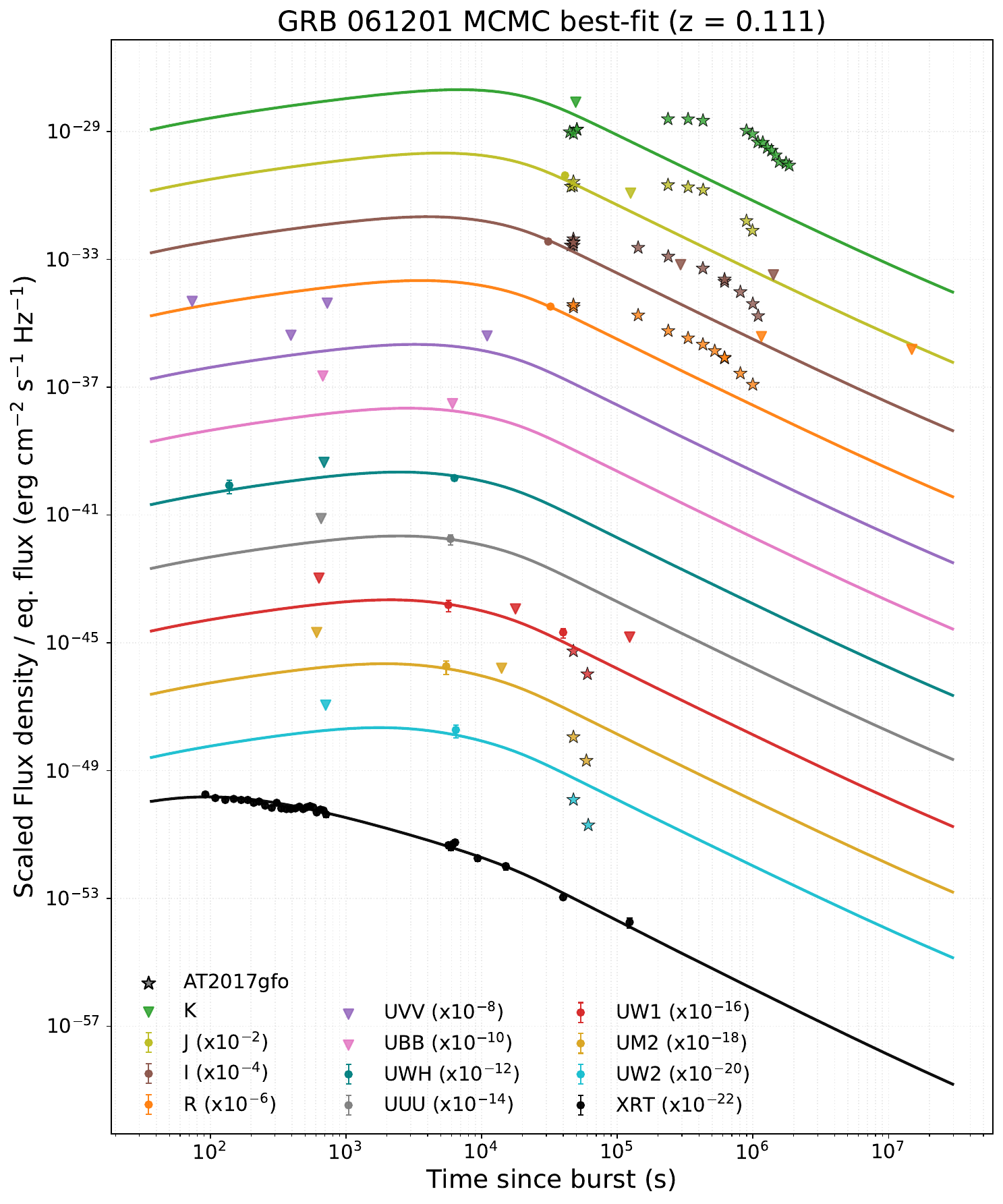}
    \hfill
    \includegraphics[width=0.48\linewidth]{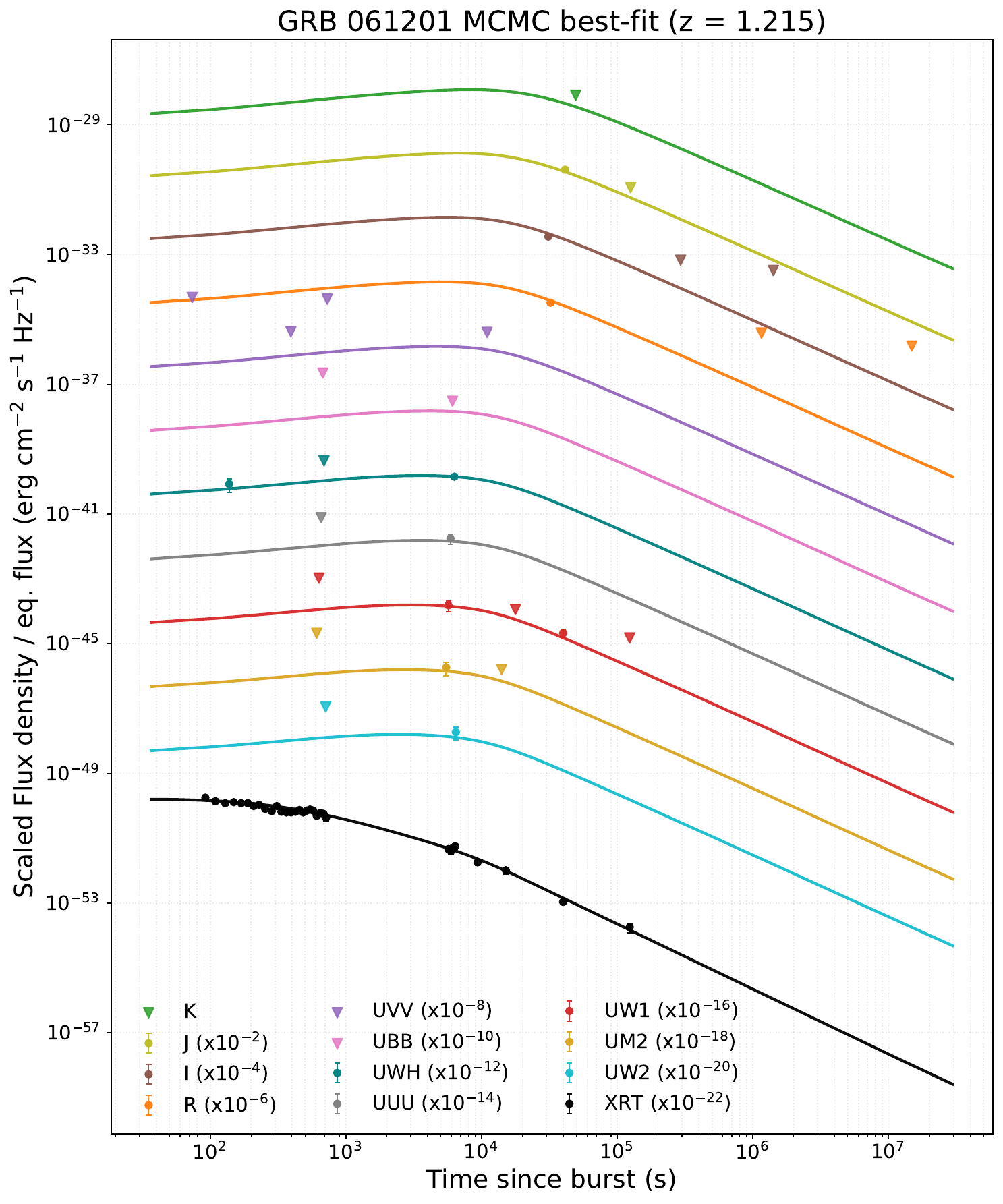}
    
    \vspace{1ex} 
    
    \makebox[0.48\linewidth][c]{(a) $z=0.111$}
    \hfill
    \makebox[0.48\linewidth][c]{(b) $z=1.2$}
    
    \caption{Multi-band afterglow modeling of GRB 061201 with \textsc{VEGASAFTERGLOW}. The left panel (a) shows the best-fitted result at $z=0.111$. For comparison, the light curve of the kilonova AT2017gfo, artificially shifted to $z=0.111$, is over-plotted (star symbols). The right panel (b) displays the best-fitted result at $z=1.2$.}
    \label{lc_fit}
\end{figure*}

The MCMC fitting results (Figure~\ref{lc_fit}) yield a minimum reduced $\chi^2/{\rm d.o.f} = 1.42$ (with 54 degrees of freedom) for the $z=0.111$ model, compared to a significantly improved reduced $\chi^2/{\rm d.o.f} = 1.18$ for the $z=1.2$ scenario. To formally quantify the model preference, we utilize the Akaike Information Criterion (AIC). The analysis returns ${\rm AIC} = -5492.93$ for $z=0.111$ and ${\rm AIC} = -5505.70$ for $z=1.2$. The resulting difference, $\Delta{\rm AIC} = 16.35 > 10$ \citep{burnham2002model}, constitutes strong statistical evidence against the local-origin hypothesis, strongly favoring the high-redshift solution.

Beyond pure statistical metrics, the $z=0.111$ scenario yields an unphysical parameter space. To reproduce the observed flux at such a low redshift, the model requires an anomalously high efficiency for electron acceleration. The energy fraction partitioned to electrons is driven to extreme values ($\epsilon_e \gtrsim 0.8$), implying a radiative efficiency approaching 100\%, which is difficult to reconcile with standard shock physics. Conversely, the $z=1.2$ fit significantly alleviates this tension, yielding less extreme microscopic energy partition parameters ($\epsilon_e \sim 0.504$, $\epsilon_B \sim 0.141$). 

\begin{figure*}[t]
  \centering
  \includegraphics[width=0.7\textwidth]{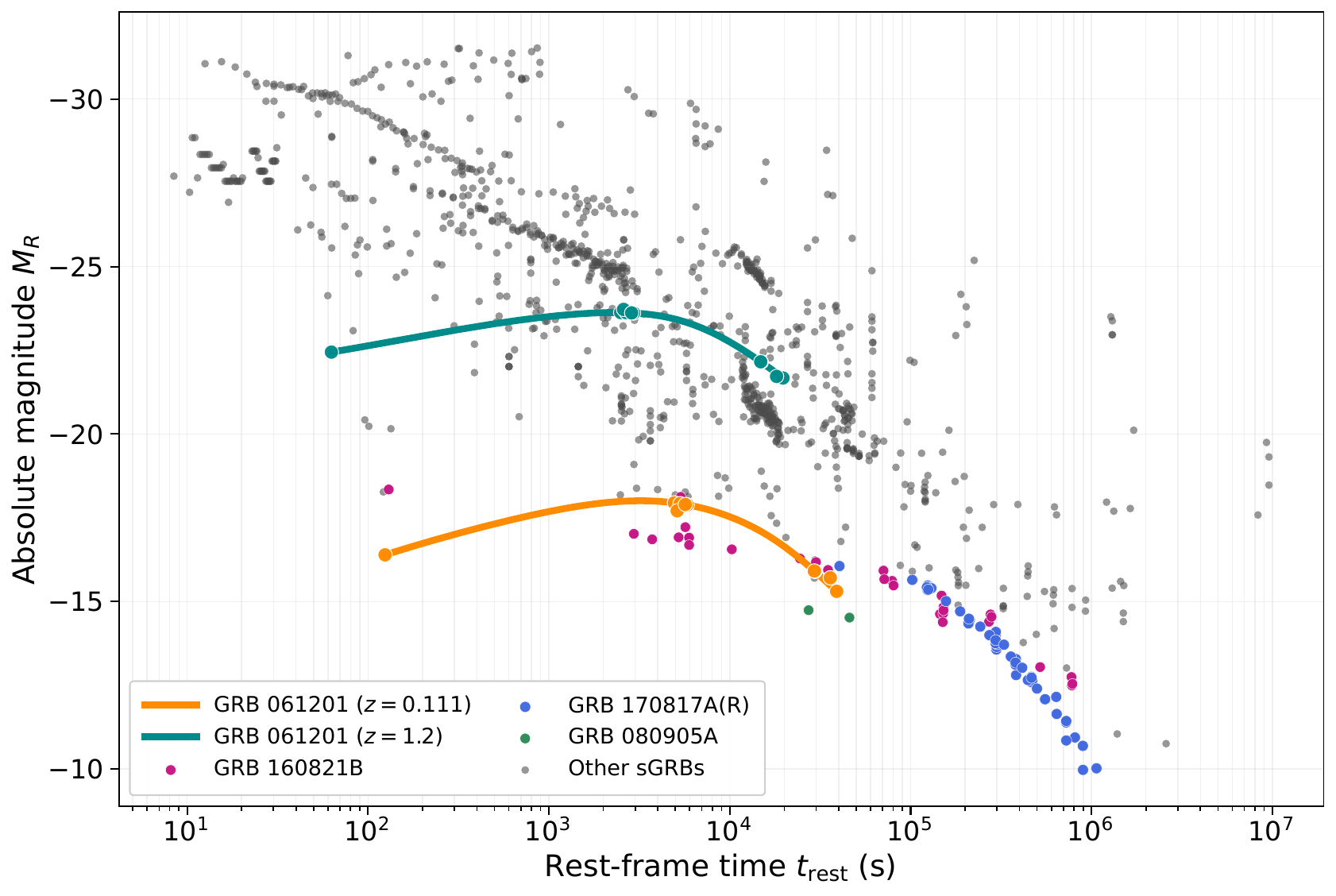}
  \caption{Light curves in different redshift of GRB 061201 in the background of other sGRBs}
  \label{fig:M-t}
\end{figure*}

To further investigate the redshift ambiguity, we compared the optical light curve of GRB 061201 with a broader population of sGRBs (Figure~\ref{fig:M-t}). We constructed a reference sample by compiling archival sGRB data, applying $k$-corrections across all bands to the rest-frame $R$ band, and converting the apparent photometry into absolute magnitudes ($M$) following the methodology of \cite{2024MNRAS.533.4023D}. These rest-frame absolute magnitude light curves ($M$--$t_{\rm rest}$) serve as our background distribution. We then overlaid the evolutionary tracks of GRB 061201 onto this parameter space under the two assumed redshifts: $z=0.111$ and $z=1.2$. As illustrated in Figure~\ref{fig:M-t}, the $z=1.2$ scenario is more consistent with the bulk demographic of the sGRB distribution. Nevertheless, we note that a $z=0.111$ origin cannot be definitively ruled out based solely on this metric, as a small number of intrinsically faint bursts still fall below its low-redshift trajectory. It is noteworthy that GRB 080905A, which lies below the $z=0.111$ curve of GRB 061201 in our $M$--$t$ diagram, shares a similar historical ambiguity. Although GRB 080905A is widely associated with a host at $z=0.1218$, high-resolution imaging with the \textit{HST}/F160W filter revealed a very faint galaxy ($m_{\rm AB} \sim 26$~mag) within $1^{\prime\prime}$ of the optical afterglow position\citep{2010MNRAS.408..383R,2021ApJ...923...38N}. This discovery, as noted by \citet{2013ApJ...776...18F}, suggests that this faint source could be the actual host at higher redshift ($z > 1$). If this high-redshift interpretation for GRB 080905A were to be confirmed, GRB 160821B would become the only burst in our sample located below the $z=0.111$ track of GRB 061201.

\subsection{Kilonova and the Merger Rate Constraints}
To rule out a local origin, we investigated the expected kilonova signature \citep{1998ApJ...507L..59L, 2010MNRAS.406.2650M, 2013ApJ...775...18B, 2017Natur.551...67P}. We shifted the light curves of the kilonova AT2017gfo\citep{2017Sci...358.1565E, 2017Sci...358.1570D, 2017ApJ...848L..27T, 2017Natur.551...67P, 2017Sci...358.1559K, 2017Natur.551...75S, 2017ApJ...851L..21V} to $z=0.111$ for a direct comparison with the observations of GRB 061201 (Figure~\ref{lc_fit}). We find that if GRB 061201 were accompanied by an AT2017gfo-like kilonova at $z=0.111$, the associated late-time emission in the optical and near-infrared ($I$- and $J$-bands) would have exceeded our observational upper limits (Figure~\ref{lc_fit}). The non-detection of such a signal provides independent, strong evidence that disfavors the $z=0.111$ local-kilonova scenario.

Furthermore, the assumed redshift of GRB 061201 significantly affects the global estimation of the binary neutron star merger rate. If GRB 061201 were a local event at $z=0.111$ with the correspondingly inferred narrow jet opening angle ($\sim 1^\circ$), the beaming-corrected merger rate derived from sGRBs would be driven to anomalously high values, reaching $1426^{+1042}_{-640}\ \mathrm{Gpc}^{-3}\ \mathrm{yr}^{-1}$. The contribution from GRB 061201 alone would account for $964^{+1002}_{-424}\ \mathrm{Gpc}^{-3}\ \mathrm{yr}^{-1}$, severely biasing the entire volumetric rate \citep{jin2026neutron}.

In contrast, the merger rate inferred from gravitational wave (GW) observations is estimated to be $165^{+183}_{-98}\ \mathrm{Gpc}^{-3}\ \mathrm{yr}^{-1}$. Compared to GRB-based population studies, GW detections provide an independent physical baseline that is largely free from beaming systematics. As an essentially isotropic messenger, gravitational radiation allows for the direct extraction of luminosity distance and component masses, circumventing the substantial uncertainties introduced by highly collimated jet geometries and complex host-galaxy redshift associations. 

Consequently, a low-redshift origin for GRB 061201 would induce severe tension between the electromagnetically inferred merger rate and the GW baseline \citep{jin2026neutron}. Recognizing this substantial rate divergence provides strong astrophysical evidence against the $z=0.111$ hypothesis. By adopting the high-redshift solution, the severe tension between the two methodologies is largely mitigated. While a mild residual discrepancy may persist---potentially hinting at deeper complexities in the intrinsic progenitor channels of sGRBs---a detailed exploration of this broader population issue is beyond the scope of this paper. Ultimately, synthesizing these multi-messenger constraints with our prior spatial and energetic diagnostics, the $z=1.2$ high-redshift origin emerges as the most self-consistent physical framework for GRB 061201.

\section{Conclusion}

Using deep JWST and archival HST imaging, we have identified a new candidate host galaxy (G2) for the short-hard GRB 061201, which is closer to the archival VLT afterglow position. Broadband photometry and subsequent \texttt{EAZY} SED modeling yield a best-fitted photometric redshift of $z_{\rm best} = 1.2$, with a 95\% highest-posterior-density credible interval of $z \in [1.14, 1.47]$. 

Comprehensive multi-band modeling indicates that G2 possesses fundamental macroscopic properties (stellar mass, age) and a physical offset that are highly representative of established sGRB host environments. While the chance-coincidence probability for G2 ($P_{\rm cc} = 0.18$) slightly exceeds conservative thresholds adopted in earlier, shallower surveys, it is well consistent with a plausible physical association once the exceptional depth and elevated source density of JWST imaging are appropriately accounted for. 

To ensure the robustness of this association, we investigated the deep localization field. We robustly exclude an alternative, fainter nearby source (G3) due to an apparent high-redshift ($z > 3$) photometric color break. With the host most plausibly associated with G2, we demonstrate that this high-redshift ($z=1.2$) solution naturally resolves the significant physical tensions that affect local-origin ($z=0.111$) hypotheses. 

Through independent and rigorous physical diagnostics---including rest-frame light curve tracking, beaming-corrected energetics, deep near-infrared kilonova limits, MCMC afterglow parameter modeling, and global binary neutron star merger rate constraints---we systematically disfavor the local-origin scenario. Taken together, all lines of evidence consistently demonstrate that GRB 061201 originated from a moderately high-redshift host galaxy that evaded detection in earlier, shallower surveys. This finding effectively mitigates the severe physical and demographic anomalies implied by the previously proposed local association. We caution, however, that deep spectroscopic observations of G2 are ultimately required to definitively confirm this photometric redshift and firmly cement the astrophysical framework presented here.

\begin{acknowledgments}
We are deeply grateful to Prof. Yi-Zhong Fan for his stimulating discussion. 
This work is supported by the Natural Science Foundation of China ((grant Nos. 12225305, 12473049, 12321003, and 12233011), the National Key R\&D Program of China (grant Nos. 2024YFA1611704 and 2024YFA1611700), the Strategic Priority Research Program of the Chinese Academy of Sciences (grant No. XDB0550400). 
J. R. is Supported by the Postdoctoral Fellowship Program and China Postdoctoral Science Foundation (grant No.~BX20250160), the Jiangsu Funding Program for Excellent Postdoctoral Talent (grant No.~2025ZB272), 
and the General Fund of the China Postdoctoral Science Foundation (grant No.~2024M763530). 
Y.W. is supported by the Jiangsu Funding Program for Excellent Postdoctoral Talent (grant 2024ZB110), the Postdoctoral Fellowship Program (grant GZC20241916), and the General Fund (grant 2024M763531) of the China Postdoctoral Science Foundation. 
H.Z. is funded by Basic Research Program of Jiangsu (No. BK20251707) and the Postdoctoral Innovation Talents Support Program (No. BX20250159). Y.M.Z is funded by Jiangsu Funding Program for Excellent Postdoctoral Talent (grant 2024ZB178).

This work is based on observations made with the NASA/ESA Hubble Space Telescope (HST) and the NASA/ESA/CSA James Webb Space Telescope (JWST). The data were obtained from the Mikulski Archive for Space Telescopes (MAST) at the Space Telescope Science Institute, which is operated by the Association of Universities for Research in Astronomy, Inc., under NASA contract NAS 5-03127 for JWST, and NAS 5-26555 for HST. This research also made use of data based on observations collected at the European Southern Observatory under ESO programme 078.D-0809(G), 079.D-0909(G), 177.A-0591(J), and based on observations obtained at the Southern Astrophysical Research (SOAR) telescope, with OBSIDs soar.osiris.\allowbreak{}20061202T020808Z, soar.osiris.\allowbreak{}20061203T003010Z, and soar.osiris.\allowbreak{}20061202T051656Z. The HST, JWST, and MAST data are available at \dataset[10.17909/m8nb-gq52]{https://doi.org/10.17909/m8nb-gq52}.
\end{acknowledgments}

\software{
    \textsf{Astropy} \citep{2013A&A...558A..33A, 2018AJ....156..123A, 2022ApJ...935..167A}, 
    \textsf{Bagpipe} \citep{2018MNRAS.480.4379C},
    \textsf{EAZY} \citep{2008ApJ...686.1503B}, 
    \textsf{Emcee} \citep{2013PASP..125..306F},
    \textsf{Matplotlib} \citep{2007CSE.....9...90H}, 
    \textsf{NumPy} \citep{2020Natur.585..357H}, 
    \textsf{SAOImage DS9} \citep{2003ASPC..295..489J},
    \textsf{SciPy} \citep{2020NatMe..17..261V}, 
    \textsf{SExtractor} \citep{1996A&AS..117..393B},
    \textsf{Starlink} \citep[Gaia;][]{2014ASPC..485..391C},
    \textsf{VegasAfterglow} \citep{2026JHEAp..5000490W}
}

\bibliographystyle{aasjournalv7} 
\bibliography{ref}               

\end{document}